%% file: ANA-HDBS-2019-23-PAPER.tex
\documentclass[cernpreprint, atlasdraft=false, texlive=2023, USenglish, texmf, orcidlogo]{atlasdoc}

\usepackage{atlaspackage}
\usepackage{atlasbiblatex}
 
\usepackage{atlasphysics}

\graphicspath{{logos/}{figures/}}
 
\usepackage{ANA-HDBS-2019-23-PAPER-defs}

\AtlasTitle{Anomaly detection search for new resonances decaying into a Higgs boson and a generic new particle $X$ in hadronic final states using $\sqrt{s} = 13$~TeV $pp$ collisions with the ATLAS detector}

\AtlasAbstract{
A search is presented for a heavy resonance $Y$ decaying into a Standard Model Higgs boson $H$ and a new particle $X$ in a fully hadronic final state.
The full Large Hadron Collider Run~2 dataset of proton--proton collisions at $\sqrt{s}= 13$~TeV collected by the ATLAS detector from 2015 to 2018 is used, and corresponds to an integrated luminosity of 139 fb$^{-1}$.
The search targets the high $Y$-mass region, where the $H$ and $X$ have a significant Lorentz boost in the laboratory frame.
A novel application of anomaly detection is used to define a general signal region, where events are selected solely because of their incompatibility with a learned background-only model.
It is constructed using a jet-level tagger for signal-model-independent selection of the boosted $X$ particle, representing the first application of fully unsupervised machine learning to an ATLAS analysis.
Two additional signal regions are implemented to target a benchmark $X$ decay into two quarks, covering topologies where the $X$ is reconstructed as either a single large-radius jet or two small-radius jets.
The analysis selects Higgs boson decays into $b\bar{b}$, and a dedicated neural-network-based tagger provides sensitivity to the boosted heavy-flavor topology.
No significant excess of data over the expected background is observed, and the results are presented as upper limits on the production cross section $\sigma(pp \rightarrow Y \rightarrow XH \rightarrow q\bar{q}b\bar{b}$) for signals with $m_Y$ between 1.5 and 6~TeV and $m_X$ between 65 and 3000~GeV.
}

\AtlasRefCode{HDBS-2019-23}

\PreprintIdNumber{CERN-EP-2023-045}

\AtlasJournalRef{Phys. Rev. D 108 (2023) 052009}
\AtlasDOI{10.1103/PhysRevD.108.052009}

\hypersetup{pdftitle={ATLAS document},pdfauthor={The ATLAS Collaboration}}
 
\begin{document}
 
\maketitle

\section{Introduction}
\label{sec:intro}
 
The Standard Model (SM) provides a framework for understanding fundamental particles and interactions that has been remarkably predictive of experimental results over several decades.
The discovery of the Higgs boson in 2012~\cite{HIGG-2012-27,CMS-HIG-12-028} completed the set of particles predicted by the SM.
 
The sensitivity of the Higgs boson mass to radiative corrections implies either extreme fine-tuning in the SM or the existence of new particles at an energy scale not far above the Higgs boson mass.
This theoretical motivation, coupled with the existing experimental mass reach of the Large Hadron Collider (LHC) at CERN, motivates searches for new particles with $\mathcal{O}(\TeV)$ masses.
Because the Higgs boson couples more strongly to heavier particles, it is natural to expect that these new heavy particles may have decays to a Higgs boson.
 
A search is presented here for a new \TeV-scale narrow-width particle \textit{Y}, which decays into a Standard Model Higgs boson $H$ and a new particle \textit{X} with a mass near the weak scale.
A fully hadronic final state is targeted for both particles.
Tagging of the boosted Higgs decay into two $b$-quarks (\Hbb~tagging) enhances a signal by using the highest branching fraction decay of the Higgs boson.
A novel jet-level implementation of \textit{anomaly detection} implemented via an unsupervised machine-learning architecture is used to select \textit{X} particles solely because of their incompatibility with the expected SM background.
 
The application of anomaly detection to collider searches is a rapidly growing effort in the high-energy physics community (see Refs.~\cite{Kasieczka_2021, Aarrestad_2022, https://doi.org/10.48550/arxiv.2112.03769} for overviews).
The lack of evidence for new interactions and particles since the Higgs boson's discovery has motivated the execution of generic searches to complement the existing rigorous, model-dependent analysis program.
Machine learning provides an excellent framework for the construction of tools that can isolate events in data solely because of their incompatibility with a background-only hypothesis.
Building a tool to perform model-independent classification of collision events involves training on data events, and therefore requires the ability to cope with a lack of labels indicating whether inputs are signal or background.
This distinguishes the typical supervised classification problem, where all inputs are labeled with a known origin, from the anomaly detection approach, which makes use of unsupervised (no input labels) or weakly supervised (noisy labels) training.
A selection tool that is constructed in this way has trade-offs: while it broadens sensitivity to many classes of new physics with a single classifier, it is expected that sensitivity to any specific model will be better with a dedicated supervised approach.

\begin{figure}[hbpt]
\centering
\includegraphics[width=0.3\textwidth]{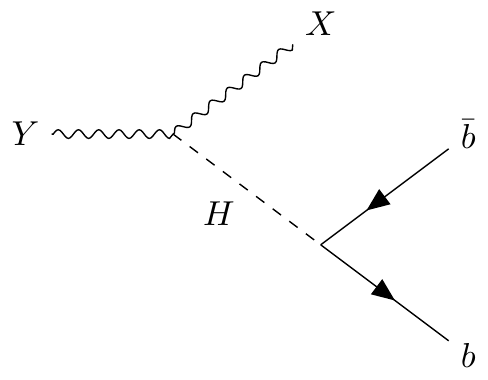}
\caption{\label{fig:feynmandiag} Feynman diagram of the target signal process, where the \textit{Y}  is produced in the initial $pp$ collision and decays to a fully hadronic final state via a SM Higgs boson $H \rightarrow b\bar{b}$ and a new particle \textit{X} . The only assumption applied to the \textit{X}  decay is that it yields a hadronic final state.}
\end{figure}

A Feynman diagram for the $Y \rightarrow XH$ signal process is shown in Figure~\ref{fig:feynmandiag}, where the \textit{X} can have a variety of hadronic decays.
A nominal benchmark decay into two light quarks, $X\rightarrow q\bar{q}$, is generated to provide an interpretation framework for the results.
The generic selection of the \textit{X} is not strongly dependent on its mass, so the analysis is sensitive to \textit{X}  masses spanning several orders of magnitude, from $\mathcal{O}(10)$~\GeV\ to $\mathcal{O}(1)$~\TeV.
The masses of the parent and daughter particles yield a kinematic scenario where the final-state particles are highly Lorentz-boosted in the laboratory frame, motivating a \enquote{boosted} reconstruction method using large-radius (large-$R$) jets and jet substructure observables to distinguish the boson decay products.
An orthogonal \enquote{resolved} reconstruction method is used to recover sensitivity to topologies where the \textit{X}  is less boosted and reconstructed as separate small-radius (small-$R$) jets, significantly extending the sensitive phase-space region.
Since the signal is resonant, it can be detected with a \enquote{bump hunt} on the invariant mass distribution of the reconstructed $X$ and $H$ candidates.
 
Although model-independent in nature, this search is motivated by several key extensions to the Standard Model which predict heavy diboson resonances.
Examples of such theories include extended gauge symmetry models~\cite{extendedgauge}, warped extra dimensions~\cite{warpeddim_1,warpeddim_2,warpeddim_3}, or two-Higgs-doublet models~\cite{2hdm}.
Signals are generated with a simplified model based on spin-1 Heavy Vector Triplets (HVT)~\cite{hvt} which reproduces a large class of explicit beyond-the-SM (BSM) models. The HVT phenomenology provides a Lagrangian that fulfills SM symmetry constraints with an isospin SU(2) triplet formed of a neutral $Z^\prime$ and two charged $W^{\prime \pm}$ bosons.
 
This search uses the full LHC \RunTwo $\sqrt{s} = 13$~\TeV\ $pp$ dataset collected by the ATLAS detector from 2015 to 2018 and corresponding to an integrated luminosity of 139~\fb.
A search for the \yxh process was previously performed by ATLAS using 36.1~\fb\ of data, assuming $X \rightarrow q\bar{q}$, and found no significant excess for \textit{Y}  masses from 1 to 4~\TeV\ and \textit{X}  masses from 50 to 1000~\GeV~\cite{yxh_201516}.
In addition to the much larger dataset, the present analysis includes several key improvements relative to the previous search, such as a neural-network-based tagger optimized for the boosted \Hbb~topology, anomaly detection for enhanced signal model independence, and the use of two orthogonal regions to capture both boosted and resolved reconstruction of the nominal \textit{X} decay into two quarks.
The ATLAS Collaboration previously leveraged anomaly detection implemented with weakly supervised machine learning in a search for narrow new resonances in dijet events using the full \RunTwo dataset~\cite{HDBS-2018-59}.
The CMS experiment performed a multidimensional search for diboson resonances with their full \RunTwo dataset of 138~\fb\ which includes sensitivity to the signature discussed here~\cite{CMS-PAS-B2G-20-009}. A maximum local (global) significance of 3.6 (2.3) standard deviations was observed for two mild excesses of events, at resonance masses of 2.1 and 2.9~\TeV\ with $X$ masses consistent with $W$, $Z$, or Higgs bosons.
 
The analysis regions in the present search are built by selecting two large-$R$ jets, with additional criteria to reduce the background to events with Higgs and \textit{X} particles in the search regions. 
Three overlapping analysis categories are defined with different $X$ selections, namely the anomaly region and two orthogonal regions optimized for the benchmark $X\rightarrow q\bar{q}$ decay.
The degree of model-independence of the \textit{X} tagging is assessed using three simulated signals with different decays of the \textit{X} particle that lead to differing large-$R$-jet topologies, in addition to the two-light-quark final state used to generate the (\my, \mx) signal grid.
The background to the signal process consists primarily of multijet events from quantum chromodynamics (QCD) processes.
It is estimated with a fully data-driven method that incorporates reweighting based on a deep neural network (DNN) to ensure good modeling.
The final fits are made to the reconstructed invariant mass distribution of the \textit{Y} candidates in overlapping ranges of \textit{X} candidate mass to increase the sensitivity to a signal.
Since no significant deviation from the estimated background is observed, results are presented as upper limits on the cross section times branching fraction for the generic HVT process.


\section{ATLAS detector}
\label{sec:detector}
 
The ATLAS detector~\cite{PERF-2007-01} is a multipurpose particle detector with a forward-backward symmetric cylindrical
geometry and a  near $4\pi$ coverage in solid angle.\footnote{ATLAS uses a right-handed coordinate system with its origin at the nominal interaction point (IP) in the center of the detector and the $z$-axis along the beam pipe. The $x$-axis points from the IP to the center of the LHC ring, and $y$-axis points upwards. Cylindrical coordinates ($r,\phi$) are used in the transverse plane, $\phi$ being the azimuthal angle around the $z$-axis. The pseudorapidity is defined in terms of the polar angle $\theta$ as $\eta = -\ln\tan(\theta/2)$.}
The inner-detector (ID) system is immersed in a 2~T axial magnetic field and provides charged-particle tracking in the range  $|\eta| < 2.5$. The high-granularity silicon pixel detector covers the vertex region and typically provides four space-point measurements per track, the first hit normally being in the insertable B-layer installed before \RunTwo \cite{ATLAS-TDR-2010-19, PIX-2018-001}. The next detector outward is the silicon microstrip tracker, which usually provides eight measurements per track. These silicon detectors are complemented by the transition radiation tracker, which enables radially extended track reconstruction up to $|\eta| = 2.0$.
Lead/liquid-argon (LAr) sampling calorimeters provide electromagnetic (EM) energy measurements with high granularity. A steel/scintillator-tile hadron calorimeter covers the central pseudorapidity range ($|\eta|<1.7$). The endcap and forward regions are instrumented with LAr calorimeters for both the EM and hadronic energy measurements up to $|\eta|<4.9$. The muon spectrometer surrounds the calorimeters and is based on three large superconducting air-core toroidal magnets with eight coils each. The field integral of the toroids ranges between 2.0 and 6.0~Tm across most of the detector. The muon spectrometer includes a system of precision chambers for tracking and fast detectors for triggering. A two-level trigger system is used to select events. The first-level trigger is implemented in hardware and uses a subset of the detector information to accept events at a rate below $100$~kHz on average depending on the data-taking conditions. This is followed by a software-based high-level trigger (HLT) that reduces the rate of selected events to 1 kHz for offline storage. An extensive software suite \cite{ATL-SOFT-PUB-2021-001} is used in data simulation, in the reconstruction and analysis of real and simulated data, in detector operations, and in the trigger and data acquisition systems of the experiment.


\section{Data and Monte Carlo simulation}
\label{sec:datamc}
 
This search is performed with $\sqrt{s}=13$~\TeV\ $pp$ collision data collected by the ATLAS detector during \RunTwo of the LHC (2015--2018).
After the application of data quality requirements~\cite{DAPR-2018-01} to ensure that all detector components are operating normally, the dataset corresponds to an integrated luminosity of
$139.0 \pm 2.4$~\ifb~\cite{DAPR-2013-01,Avoni_2018}.
Data events used in this analysis were triggered by the presence of a high-\pt large-$R$ jet, built at trigger level using calorimeter-cell energy clusters calibrated to the hadronic scale utilizing the local cell-signal weighting method~\cite{PERF-2014-07}.
The unprescaled single-large-$R$-jet trigger with the lowest jet-\pt threshold was used in each year, corresponding to thresholds of 360~\GeV\ in 2015, 420~\GeV\ in 2016, and 440~\GeV\ in 2017 and 2018.
Further offline selection criteria are imposed (see Section~\ref{sec:selection}) to ensure that selected events are all in the kinematic region where the trigger is fully efficient.

Monte Carlo (MC) event generators were used to simulate the signal targeted by this search.
The simulated event samples include the effect of additional $pp$ interactions (\pileup) in the same or neighboring bunch crossings.
The effect is assessed by overlaying simulated minimum-bias events on each hard-scatter event and accounting for how interactions in the previous or following bunch crossing affect the detector response.
The simulated events are weighted to reproduce the pile-up distribution observed in each data-taking period.
The detector response was simulated~\cite{SOFT-2010-01} using \GEANT~\cite{Agostinelli:2002hh} and the events were reconstructed with the same software as used for data.
 
A generic HVT~\cite{evgen_HVT, hvt} 
model with simulated $qq$ scattering is used as a baseline for the \yxh signal samples.
The $Y$ resonance is assumed to have a width that is narrow compared to the detector resolution.
The generation of the $W^{\prime} \rightarrow WH$ process was modified to replace the $W$ boson with a new spin-1 boson \textit{X}  with a width of 2~\GeV\ and a 100\% branching fraction to $u\bar{d}$.
The Standard Model Higgs boson was generated with decays into $b\bar{b}$ only.
Samples with \textit{Y} resonance masses of 1--6~\TeV\ were generated using
\MGNLO[2.7.2] and 2.7.3~\cite{Alwall:2014hca} interfaced to \PYTHIA[8.2]~\cite{Sjostrand:2007gs} for parton showering and hadronization
with the \NNPDF[2.3lo] parton distribution function (PDF) set~\cite{pdfs}, at leading order in QCD, and parameter values set according to the A14 tune~\cite{tune}.
A two-dimensional grid was generated for 195 signal points defined by \textit{Y} and \textit{X} mass values, with \textit{Y} masses in the range 1.5--6~\TeV\ and \textit{X} masses in the range 65--3000~\GeV.
The $Y$ and $X$ mass values account for the kinematic constraint $\my > \mx$, and the $X$ mass boundaries are dictated by achievable sensitivity.
The signal acceptance times the efficiency has values ranging from a few percent to 40\% for the various signal region selections.
 
In addition to the \yxh signal grid, three signals with different jet topologies were used to craft and assess the degree of model independence in the $X$ anomaly-tagging procedure.
They were all generated with \PYTHIA[8] using the same tune (A14) and PDFs (\NNPDF[2.3lo]) as the \yxh signal grid.
The HVT $W' \rightarrow WZ$ configuration from Ref.~\cite{vvjj_fullrun2} was used to create generic resonance signatures of the form $A \rightarrow BC$, where the daughter particles decay into either three light quarks or two heavy-flavor quarks.
Choosing as benchmarks a parent $A$ mass of 3~\TeV\ and daughter masses of 200 and 400~\GeV\ creates approximately the same kinematic scenario as the \yxh phase space of interest, leading to boosted decay products that are captured as large-$R$ jets.
Depending on the daughter decay into either $qqq$ or $b\bar{b}$, the large-$R$ jet is produced with three-prong substructure or displaced vertices, respectively.
Lastly, the \PYTHIA[8] Hidden Valley Model A configuration was used to generate a \enquote{dark jet} signal that arises from the decay of a $Z'$ dark sector mediator into two dark quarks, where the $Z'$ has the same benchmark mass of 3~\TeV~\cite{Park:2017rfb}.
These jets contain dark matter particles that do not interact with the detector, creating a hadronization pattern that contains both visible and invisible energy.
Such jets are difficult to recognize using traditional substructure variables that focus on specific signal topologies, making them an ideal target for anomaly detection based on background-only characterization.


\section{Object selection}
\label{subsec:obj}
 
Charged-particle tracks measured to have $\pt> 500$~\MeV\ in the inner detector are used to reconstruct $pp$ collision vertices, and the one with the largest $\Sigma\pt^2$
of such tracks is chosen as the event's primary vertex.
 
Jets are built with the anti-$k_t$ algorithm~\cite{Cacciari:2008gp} in \textsc{FastJet}~\cite{Fastjet}, using two radius parameter values to cluster the constituent detector signals: $R=1.0$ for large-$R$ jets $J$, and $R=0.4$ for small-$R$ jets $j$.
Large-$R$ jets must satisfy $\pt > 200$~\GeV\ and $|\eta| < 2.0$.
They are constructed from combinations of tracks and calibrated calorimeter energy clusters known as Track-CaloClusters (TCCs)~\cite{ATL-PHYS-PUB-2017-015}.
The TCCs take advantage of both the excellent energy resolution of the ATLAS calorimeter and the better angular resolution of the ID tracking system at very high energy, where the calorimeter is less able to discern jet substructure.
This is particularly advantageous for measuring substructure in highly boosted jets, as is needed in this search to distinguish \yxh signal from multijet background.
To minimize the impact of \pileup, large-$R$ jets are trimmed~\cite{Krohn:2009th} by removing any $R=0.2$ subjets carrying less than 5\% of the large-$R$ jet's \pt.
The jet energy and mass is calibrated with a MC-based method~\cite{jetcalib}.
 
Small-$R$ jets are built from particle-flow objects~\cite{PERF-2015-09}, with improved accuracy of the jet's charged-hadron energy component through measurements of the associated charged-particle momenta in the ID.
Small-$R$ jets must have $\pt> 20$~\GeV\ and $|\eta| < 4.5$.
An additional event-level veto is applied to reject events with a jet that is likely to have come from calorimeter noise, beam-induced background, or a cosmic ray~\cite{cleaning}.
 
Leptons, while excluded from the analysis region selections, participate in an overlap removal procedure that prevents double-counting of overlapping objects.
Leptons must have $\pt > 7$~\GeV, with $\abseta < 2.47$ for electrons and $\abseta < 2.7$ for muons.
To ensure that each lepton's track originates from the primary vertex, the transverse impact parameter $d_0$, measured relative to the beam line, must have a significance $|d_0/\sigma(d_0)| < 3\,(5)$ for muons (electrons), and the longitudinal impact parameter $z_0$, from the primary vertex to the point where $d_0$ is measured, must satisfy $|z_0\sin\theta| < 0.5$~mm. No lepton isolation criteria are applied.
 
The overlap removal procedure is applied to all selected leptons and jets.
If two electrons share the same track, or the separation between their two energy clusters satisfies $|\Delta \eta| < 0.075$ and $|\Delta\phi| < 0.125$, then the lower-\pT electron is discarded.
Electrons that fall within $\Delta R = 0.02$ of a selected muon are also discarded. For electrons and nearby small-$R$ jets, the jet is removed if the separation between the electron and jet satisfies $\Delta R < 0.2$; the electron is removed if the separation satisfies $0.2 < \Delta R < 0.4$.
For muons and nearby small-$R$ jets, the jet is removed if the separation between the muon and jet satisfies $\Delta R < 0.2$ and if the jet has less than three tracks or the energy and momentum differences between the muon and the jet are small; otherwise the muon is removed if the separation satisfies $ \Delta R < 0.4$.
To prevent double-counting of energy from an electron inside a large-$R$ jet, the large-$R$ jet is removed if its separation from the electron satisfies $\Delta R < 1.0$.


\section{Variational recurrent neural network for anomalous-jet tagging}
\label{sec:vrnn}
 
The anomaly detection in this search is performed with a jet-level Anomaly Score (\SA)~\cite{akahn}.
The \SA value is given by a variational recurrent neural network (VRNN)~\cite{vrnn}, which consists of a variational autoencoder (VAE) whose latent space is updated at each time step of a recurrent neural network (RNN).
The key features of this tool as it pertains to this analysis application are provided below, with a comprehensive description provided in Ref.~\cite{akahn}.
 
The class of models that are based on autoencoders are popular for the selection of rare or outlier events in high-energy physics~\cite{Heimel_2019, Farina_2020, Bortolato_2022, Cheng_2023, Blance:2019ibf, Cerri:2018anq}.
Such an architecture is composed of two separate stages: an encoder that compresses the input into a lower-dimensional latent space through identification of its salient features, and a decoder that samples from the latent space and attempts to reproduce the input in its original dimensionality.
Minimizing the reconstruction error between input and output is a key element of the training, and allows the autoencoder to implicitly learn the underlying distribution of features in its input dataset, and thus recognize anomalous data through its inability to reconstruct such events with low error.
 
VAEs extend this concept to perform Bayesian inference via a latent space that can accommodate a distribution of encoded information.
The VRNN architecture combines the variational inference capabilities of a VAE with the sequence modeling provided by an RNN.
The inclusion of a learned, time-dependent prior distribution is an essential feature of the VRNN, and allows the modeling of complex structured sequences with high variability.
 
The VRNN is trained on large-$R$ TCC jets in the ATLAS \RunTwo dataset satisfying the criteria in Section~\ref{sec:selection} for full trigger efficiency, the jet requirements described in Section~\ref{subsec:obj}, and $\pt(J) >1.2$~\TeV.
Its architecture is the same as that described in Ref.~\cite{akahn}.
Training is performed using the \textsc{PyTorch} deep-learning library~\cite{pytorch}, and the network is updated using the Adam optimizer~\cite{kingma2017adam} with a learning rate parameter of 10$^{-5}$.
No regularization via weight decay is applied, but gradient clipping is used with a clip value of 10.
 
The input jets are modeled as a sequence of up to 20 constituent four-vectors per jet, ordered in $k_t$-splitting starting from the highest-\pt constituent.
The \pT selection is designed to chose input jets with highly boosted topologies, for which the VRNN is expected to most improve the sensitivity over the substructure-based approach, as these topologies are well-described by $k_t$-sorted sequence modeling and are also challenging to distinguish with substructure methods.
The training is conditioned on four high-level variables, namely the energy-correlator substructure variable $D_2$~\cite{Larkoski_2014,larkoski2016analytic} for two-prong sensitivity, the $N$-subjettiness ratio $\tau_{32}$~\cite{nsubjet} for three-prong sensitivity, and the two $k_t$-splitting-scale ratios $d_{12}$ and $d_{23}$~\cite{ktsplitting}.
An alignment procedure applied to each jet rescales to the same \pt, boosts to the same energy, and rotates to the same orientation in $\eta$ and $\phi$, inhibiting the VRNN from selecting jets with rare kinematic properties without considering internal constituent structure.
This input modeling is designed to reveal correlations between constituents and substructure, allowing the VRNN to distinguish jets with anomalous energy-deposition patterns from the background of homogenous jets originating from QCD processes.
 
Since the input consists solely of jets from data, no labeling scheme is used in training, distinguishing this method of unsupervised learning from traditional supervised machine learning, where the input is labeled in signal or background categories.
Of the data meeting the above criteria, 90\% is used for training, and 10\% for validation.
No blinding criterion is applied to the training data, so events that are used in training may also enter the final analysis regions.
Since the VRNN performance is consistent across a variety of signal contamination fractions in training~\cite{akahn}, and the presence of signal in the training dataset would only lower the sensitivity and cannot produce a false excess, biasing the final result is not a concern.
 
The loss function of the VRNN is composed of two terms: a reconstruction error term, to minimize differences between the decoded result and the original input, and the Kullback--Leibler (KL) divergence $D_{\text{KL}}$~\cite{Kullback59} of the encoded approximate posterior distribution from the Gaussian prior distribution.
The loss $\mathcal{L}(t)$ for each time step $t$ is given by Eq.~(\ref{eq:vrnnloss}), where $\textbf{x}(t)$ is the input constituent four-vector, $\textbf{y}(t)$ is the output constituent four-vector, $z$ is the approximate posterior, $z_t$ is the learned prior, and $\lambda$ is a constant parameter to weight the $D_{\text{KL}}$ contribution:
 
\begin{equation}
\label{eq:vrnnloss}
\mathcal{L}(t) = |\textbf{y}(t) - \textbf{x}(t)|^2 + \lambda D_{\text{KL}}(z||z_t).
\end{equation}
Here, the time step of the RNN refers to the location in the input sequence.
The learned prior $z_t$ is a function of the current time step's hidden state, and the hidden state is updated via a recurrence relation after each time step.
An overall loss $\mathcal{L}$ over the sequence is then computed by averaging the individual time-step losses over the length of the sequence.
 
The KL divergence of the encoded posterior distribution from the true posterior distribution is jointly minimized along with the loss $\mathcal{L}(t)$, ensuring that the architecture accurately describes the underlying distribution of data~\cite{kingmawelling}.
The $D_{\text{KL}}$ term has been shown to provide better discrimination between anomalous and standard jets than either the reconstruction error or the loss term as a whole.
The Anomaly Score is therefore defined in terms of the $D_{\text{KL}}$ of each constituent, averaged over the whole jet, and restricted to the range of (0, 1) via exponentiation, as shown in Eq.~(\ref{eq:as}).
The resulting variable \SA ensures that more-anomalous jets populate higher values and background populates lower values, and is defined as
 
\begin{equation}
\label{eq:as}
\SA = 1 - \text{e}^{-D_{\text{KL}}}.
\end{equation}
 
The performance of the VRNN can be assessed by applying Eq.~(\ref{eq:as}) to both background and a variety of signal models, and studying the discrimination power of \SA.
Its performance in this analysis, and the use of the Anomaly Score to select the $X$ jet, are described in the following section.


\section{Event selection}
\label{sec:selection}
 
The experimental signature of the \yxh signal contains at least two jets with high transverse momentum.
Signal-like events are selected through event-level criteria, along with jet-tagging criteria for both the $X$ and $H$ boson.
In total, the analysis is performed three times, once for each of the three signal regions (SRs).
The three analysis categories are defined using the selection criteria for the $X$ particle.
Each signal region utilizes five background estimation regions, composed of three control regions (CRs) and two validation regions (VRs), which are defined using common selections on the Higgs boson.
The definitions of these regions and motivation for the selection criteria are provided in this section.
 
The analysis uses data events passing a high-\pt-threshold single large-$R$-jet trigger. Selected events must have a primary vertex
and at least two large-$R$ jets.
Only the two highest-\pT large-$R$ jets in an event are considered further, and at least one of their masses must exceed 50~\GeV\ to remove poorly reconstructed events.
To ensure that the trigger is fully efficient, the invariant mass of the two large-$R$ jets must satisfy $m_{JJ}> 1.3$~\TeV\ and the leading large-$R$ jet, $J_1$, must have $\pt(J_1) > 500$~\GeV.
These requirements, along with a selection on the \D score defined in the next subsection, compose the analysis preselection.
 
As mentioned previously, the three analysis SRs vary in the selection criteria for the $X$ particle.
The primary SR, using the VRNN for generic $X$ tagging, is referred to as the \textit{anomaly SR}.
The other two SRs target the benchmark $X \rightarrow q\bar{q}$ decay and are thus called the \textit{two-prong SRs}.
Here the jet substructure is used to distinguish reconstruction of the $X$ particle as either a single large-$R$ jet (\textit{merged SR}) or two small-$R$ jets (\textit{resolved SR}).
They are not required to be orthogonal to the anomaly SR, but do not overlap with each other.
The analysis CRs include the high sideband (HSB) of the Higgs boson candidate mass \mH, used to generate the DNN reweighting for the background estimation procedure, and CR0, which provides the final background template for the SR.
The VRs are defined in the low sideband (LSB) of the Higgs boson candidate mass, and validate the reweighting derived from the HSBs.
The full analysis flow is shown in Figure~\ref{fig:analysisflow}, with selection details provided below.
 
\begin{figure}[h!]
\centering
\includegraphics[width=0.8\textwidth]{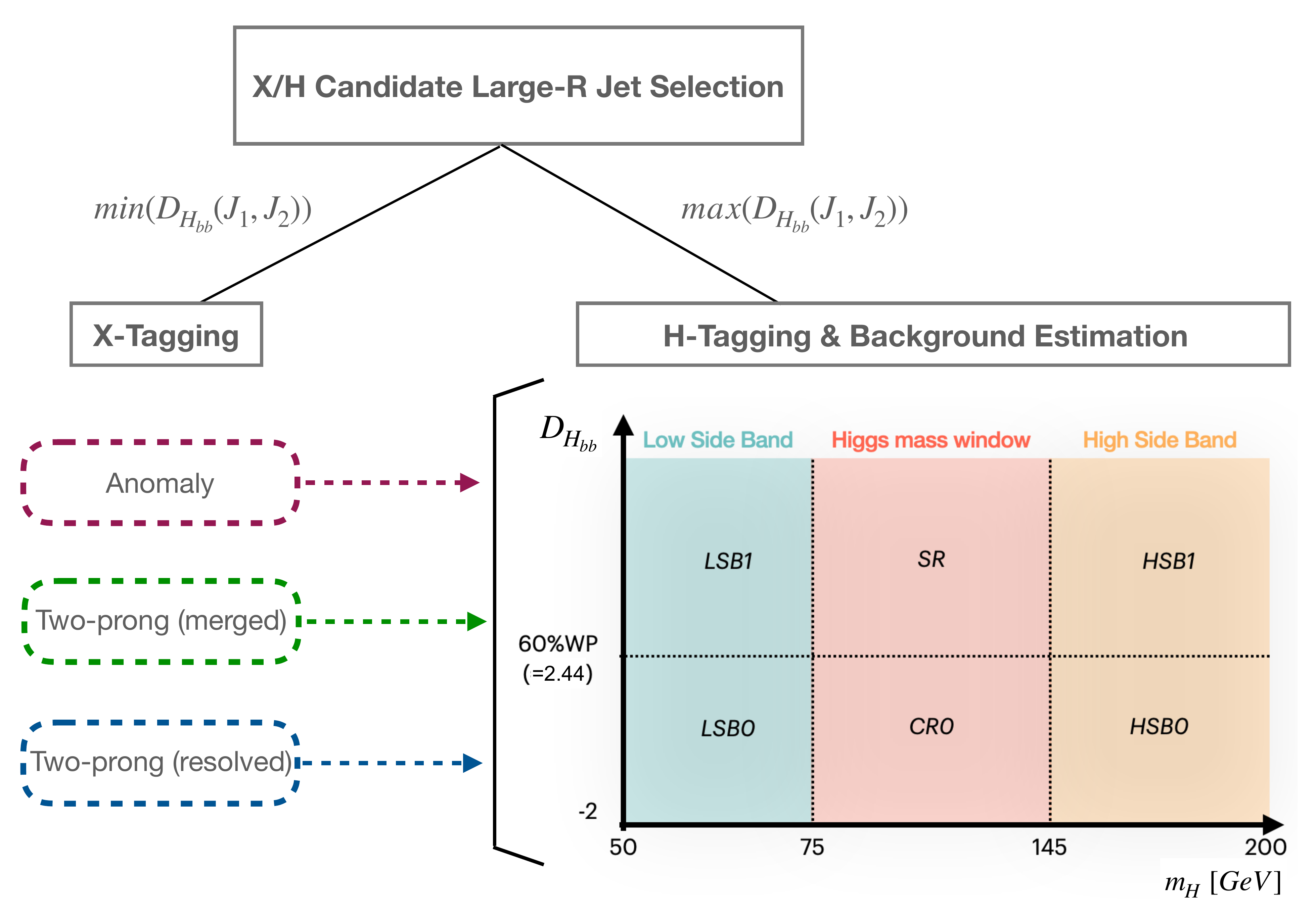}
\caption{\label{fig:analysisflow} Illustration of the selection flow, after preselection, and the analysis regions of the \yxh search. The first step in resolving the ambiguity in assigning the two large-$R$ jets to $X$ and $H$ candidates dictates which jet is used to make the first selection on either \dtrk or the Anomaly Score of the $X$ candidate. This sorts the preselection events into three separate categories, namely anomaly, two-prong merged, and two-prong resolved.  Each region has its own background estimation and validation regions. The CRs (CR0, HSB0, HSB1), VRs (LSB0, LSB1), and SR are defined in the same way for all three analyses using the $H$ candidate mass (\mH) and \D score. The large-$R$ jet $J_H$ is required to have $\D > -2$.}
\end{figure}
 
\subsection{$X$/$H$ ambiguity resolution}
 
Since the SRs are constructed by placing requirements on the different properties of the $X$ and $H$ large-$R$ jets, an ambiguity must be resolved to determine which of the two $J$ in the event is more likely to be the Higgs boson and thus subject to the Higgs boson selection criteria.
This is done by using a classifier based on a neural network (NN) to separate bosons decaying into $b\bar{b}$ from top-quark and QCD jets~\cite{xbb}.
The tagger is trained on the large-$R$ jet \pt and $\eta$, along with the subjet flavor-tagging score DL1r~\cite{FTAG-2018-01} for up to three track-subjets constructed with a jet-\pt-dependent radius parameter~\cite{vrjets}.
The tagger version used here includes a reweighting of all training inputs to have the same \pt and $\eta$ distributions, to minimize biasing of the tagger towards high-\pt or central jets respectively.
The outputs of the NN are three classification scores corresponding to the probability for the jet to have originated from a Higgs boson ($P_{\text{Higgs}}$), top quark ($P_{\text{top}}$), or multijet process ($P_{\text{multijet}}$), and these are subsequently combined into the jet-level discriminant \D:
\begin{equation}
\label{eq:dhbb}
D_{H_{bb}} = \ln\frac{P_{\text{Higgs}}}{\ftop\cdot P_{\text{top}} + (1-\ftop)\cdot P_{\text{multijet}}}.
\end{equation}
In Eq.~(\ref{eq:dhbb}), \ftop determines the weight assigned to the top-quark background shape in the final discriminant, and its chosen value of 0.25 is based on signal-to-noise optimization studies.
 
To perform the large-$R$ jet ambiguity resolution, \D is computed for both $J$ in the event.
The jet with the larger value of \D is labeled as the Higgs boson candidate ($J_H$), and the other $J$ is by default the $X$ candidate ($J_X$), thus determining which jet is subject to further $H$ and $X$ selection.
This procedure has an accuracy of over 90\% in the highly boosted region of the signal grid ($\mx/\my < 0.5$), decreasing to no less than 75\% for less boosted mass points where the boosted \Hbb tagging is less efficient.
The resulting distribution of \D for the $J$ chosen as the Higgs boson candidate, for both data and representative \yxh signals at preselection, is shown in Figure~\ref{fig:vars_dhbb}.
Additionally, a preselection of $\D > -2$ is applied to remove events that are determined to be not \Hbb-like, thus ensuring the data-driven background estimation focuses on a phase-space region that is close to that of the signal.
 
\begin{figure}[hbpt]
\centering
\includegraphics[width=0.45\textwidth]{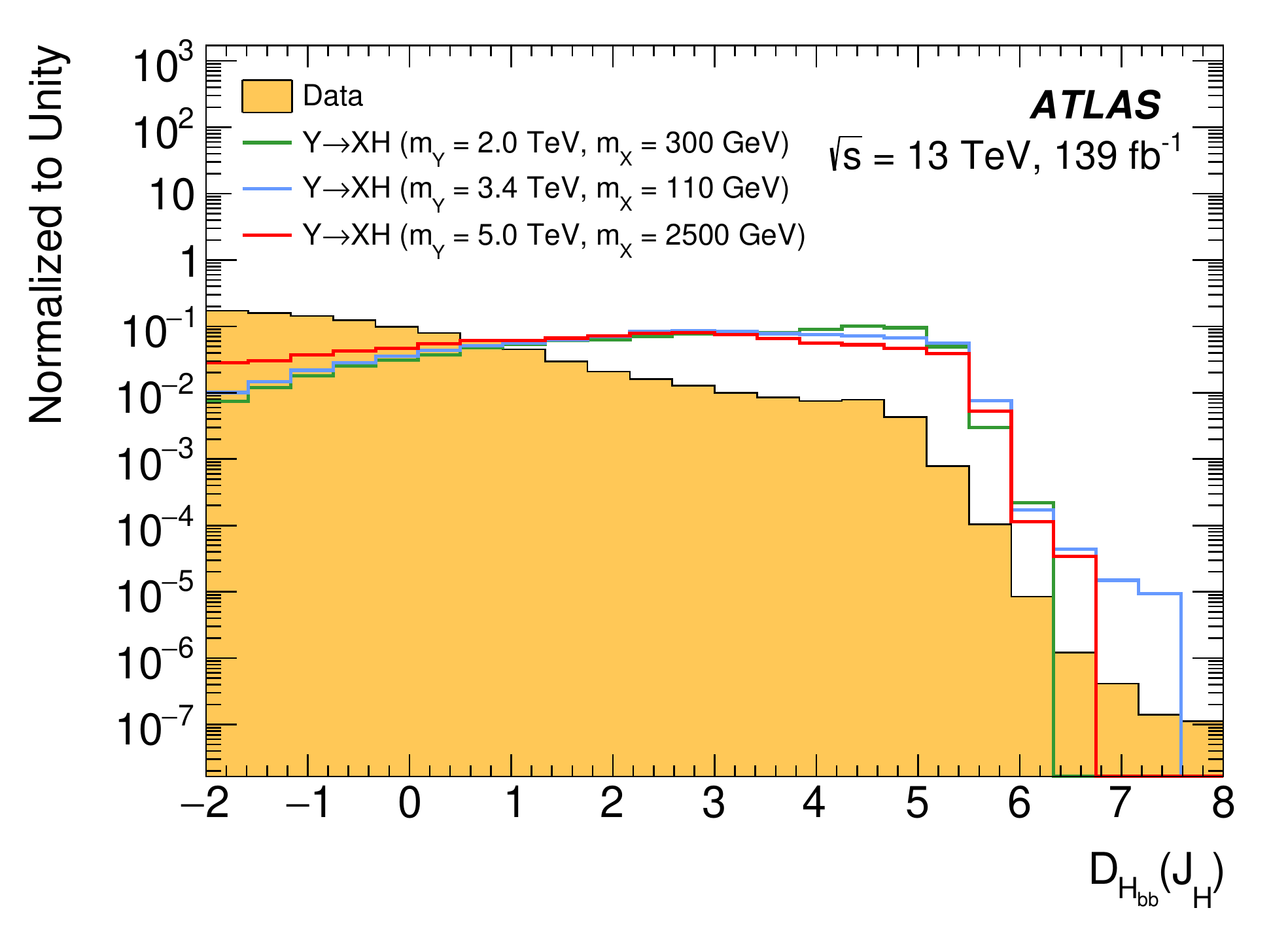}
\caption{\label{fig:vars_dhbb} Distributions of the $J_H$ candidate \D score in data after preselection requirements are applied. Also shown are three \yxh simulated signals, labeled by the masses of the $Y$ and $X$ particles. Requiring $\D > 2.44$ defines a working point that is 60\% efficient for the selection of the boosted $H\rightarrow b\bar{b}$ topology across the full \pt range. All distributions are normalized to unity.}
\end{figure}
 
\subsection{$X$ selection}
 
After the $X$ candidate's $J$ is determined, a region-dependent selection is applied to enhance the anomalous-jet content in the anomaly, two-prong merged, and two-prong resolved regions.
The anomaly region is defined using the Anomaly Score value \SA of the $X$ jet as determined by the VRNN.
Figure~\ref{fig:as} shows the resulting distribution of $X$ candidate \SA values after the analysis preselection is applied, comparing data with three different \yxh baseline signal samples as well as the three signals with alternative $X$ decays.
The absence of signal information in the construction of the score means that it is not expected to outperform analytic or supervised methods for any single signal model.
However, using the VRNN approach to characterize the background distribution provides broad sensitivity to a variety of signals that differ from jets originating from QCD processes.
The most notable difference in \SA shape is found for dark jets, which have a substructure that is not well-characterized by existing variables.
The dependence of the sensitivity on kinematics is also relevant to performance, as the \SA distribution for the highly resolved point ($\my = 5000$~\GeV, $\mx = 2500$~\GeV) populates lower, and thus less anomalous, values than the data, making this signal indistinguishable from background in the anomaly detection approach.
A selection $\SA > 0.5$ is chosen for the $J_X$ candidate, and provides modest but comparable discrimination power in data for the disparate signal-jet topologies considered in the full expected \pt range of the $X$ candidate jets. 
This enhances the signal-to-background ratio for the two boosted points, ($\my = 2000$~\GeV, $\mx = 300$~\GeV) and ($\my = 3400$~\GeV, $\mx = 110$~\GeV), by approximately 25\% relative to the inclusive selection.
 
\begin{figure}[hbpt]
\centering
\includegraphics[width=0.45\textwidth]{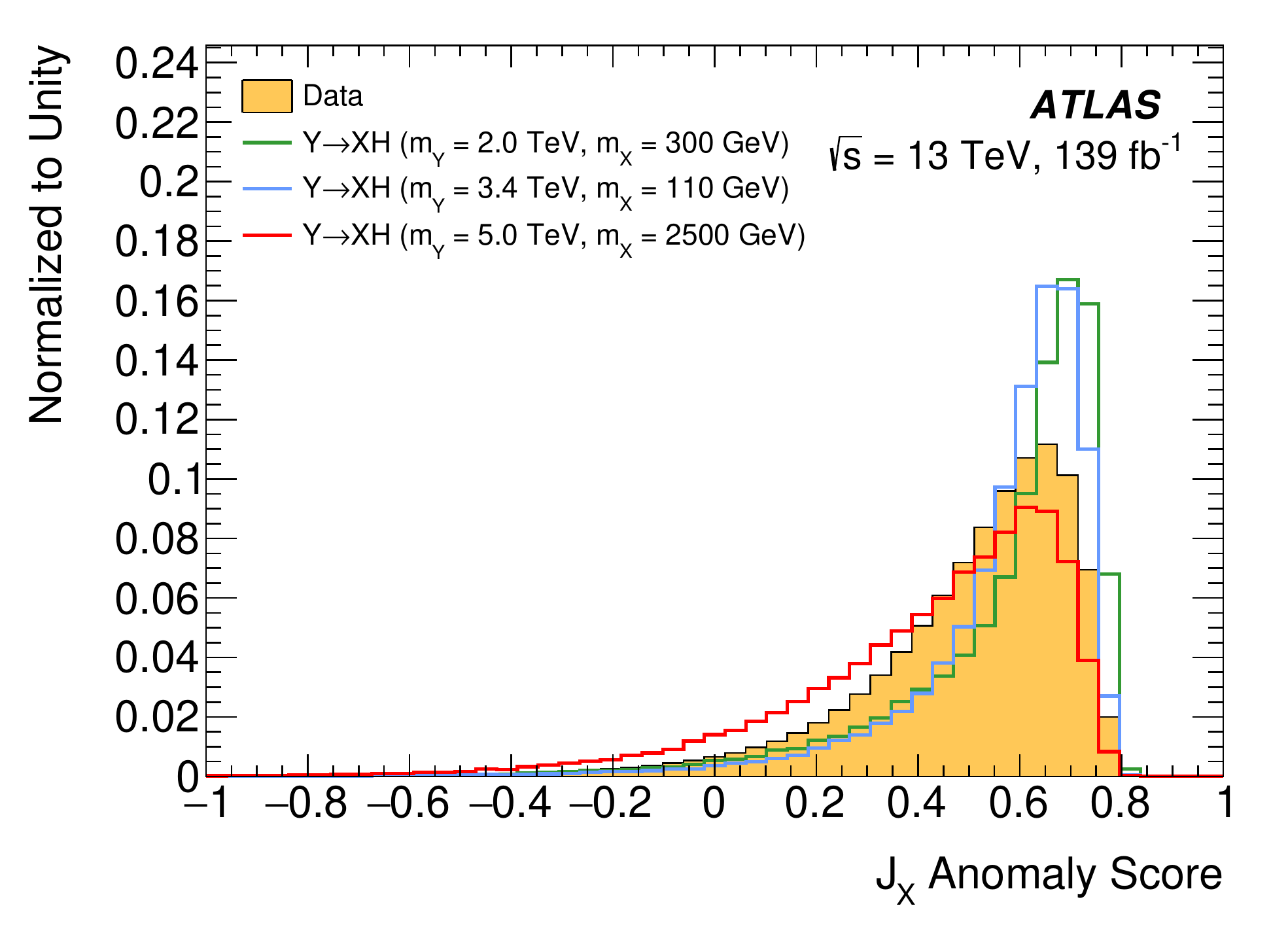}
\includegraphics[width=0.45\textwidth]{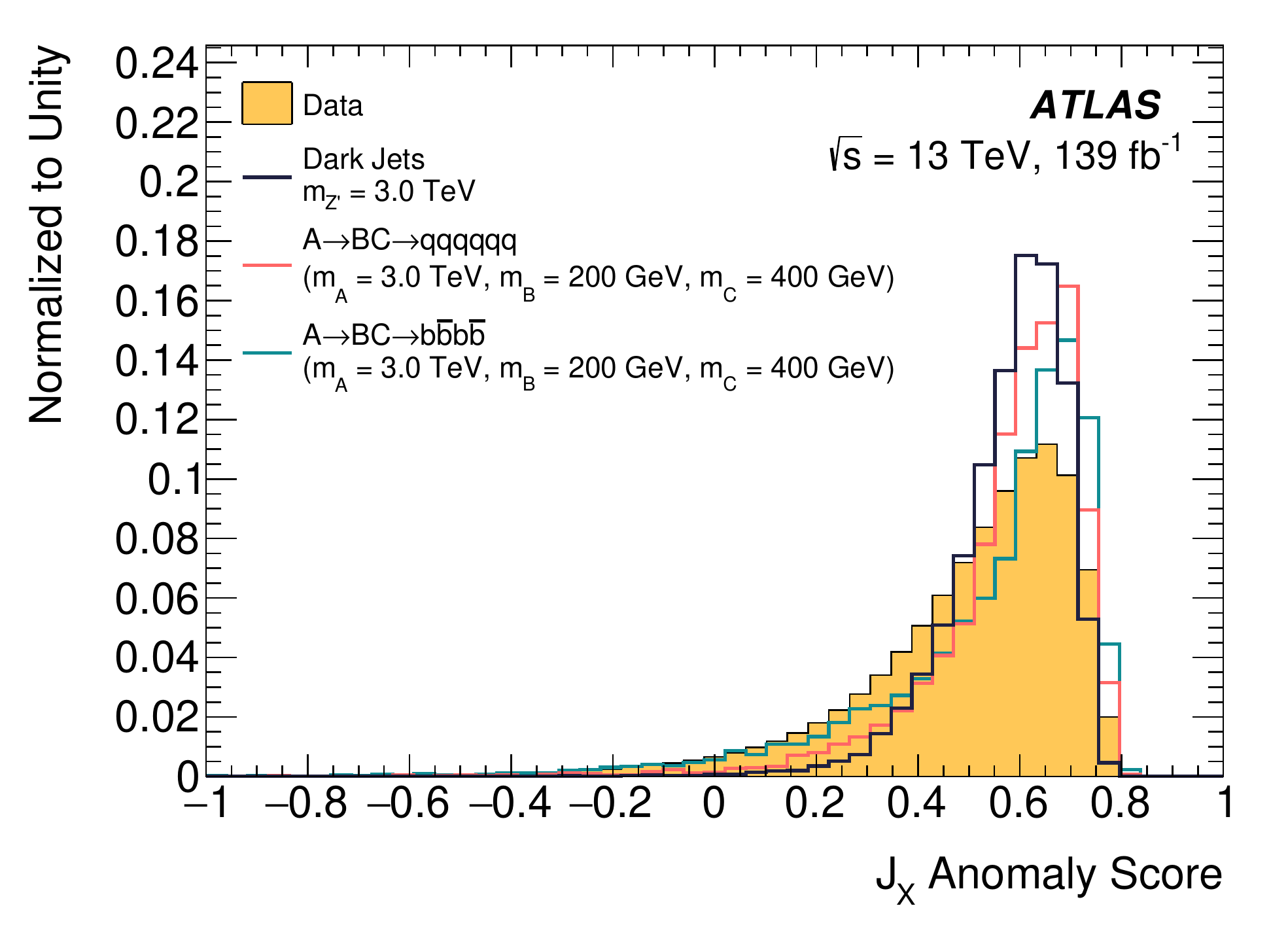}
\caption{\label{fig:as} Distributions of the $J_X$ candidate Anomaly Score (\SA) in data after preselection requirements are applied. Also shown are three \yxh simulated signals (left), labeled by the masses of the $Y$ and $X$ particles, and the three additional signals with alternative $X$ decay hypotheses, namely heavy-flavor, three-prong, and dark jet (right). A selection $\SA > 0.5$ applied to $J_X$ defines the anomaly analysis category. All distributions are normalized to unity.}
\end{figure}
 
Regions focusing on the benchmark $X\rightarrow q\bar{q}$ decay are used to supplement the anomaly detection analysis and interpret the results.
The merged region is defined by applying a selection $\dtrk < 1.2$, where \dtrk is the same as $D_2$ but computed using only tracks associated with the jet.
The choice of using only tracks in the $D_2$ calculation is motivated by the ability to propagate track-only uncertainties into the final signal efficiency.
Detailed comparisons of \dtrk with the standard $D_2$ variable at analysis preselection have shown \dtrk and $D_2$ to be highly compatible, ensuring its sensitivity to two-prong signals with respect to multijet background.
Because jets with two-prong substructure have lower values of \dtrk, this selection enhances the presence of fully merged decays where the $X$ decay products are well-contained by a single $J$.
Since the efficiency of substructure variables is mainly driven by the boost of the large-$R$ jet and therefore its \pt, the
optimal cut value of 1.2 was chosen so as to maximize signal-vs-background discrimination across the \pt range of the $X$ candidate, combining all mass points in the \yxh grid.
Distributions of \dtrk for $J_X$ are shown in Figure~\ref{fig:X_vars} for both data and representative \yxh signals at preselection.
 
Due to kinematic constraints driven by the \mx/\my ratio, events that fail the merged selection typically also show poor reconstruction of the $X$ mass by the large-$R$ jet.
In these events, the large-$R$ jet does not sufficiently contain all of the $X$ decay products, leading to an underestimation of the $X$ particle mass.
To achieve better sensitivity in this regime, a resolved selection is defined by requiring the \dtrk value of $J_X$ to be greater than 1.2, corresponding to a less boosted $X$ particle which is more appropriately reconstructed using two small-$R$ jets.
A dedicated algorithm matches two small-$R$ jets to the $X$ by requiring at least four $j$ in the event, discarding the two with the smallest $\Delta R$ to the Higgs boson candidate $J_H$, and selecting the two remaining small-$R$ jets with the highest \pt.
The mass of the $Y$ resonance is then computed using the Higgs boson candidate large-$R$ jet and the two small-$R$ jets matched to the $X$.
Additional requirements of $|\Delta y_{j1,j2}| < 2.5$ and a \pT imbalance $\pT^{\text{bal}} < 0.8$ are applied to aid accurate resolved $X$ reconstruction; here $\Delta y$ refers to separation in rapidity and $\pT^{\text{bal}}$ is defined as the difference of the $X$ particle's small-$R$ jet transverse momenta divided by their sum.
The resolved region improves the signal efficiency for points with higher \mx/\my (corresponding to less boost for the $X$ candidate decay) by a factor of approximately five relative to the boosted selection.
 
\begin{figure}[hbpt]
\centering
\includegraphics[width=0.45\textwidth]{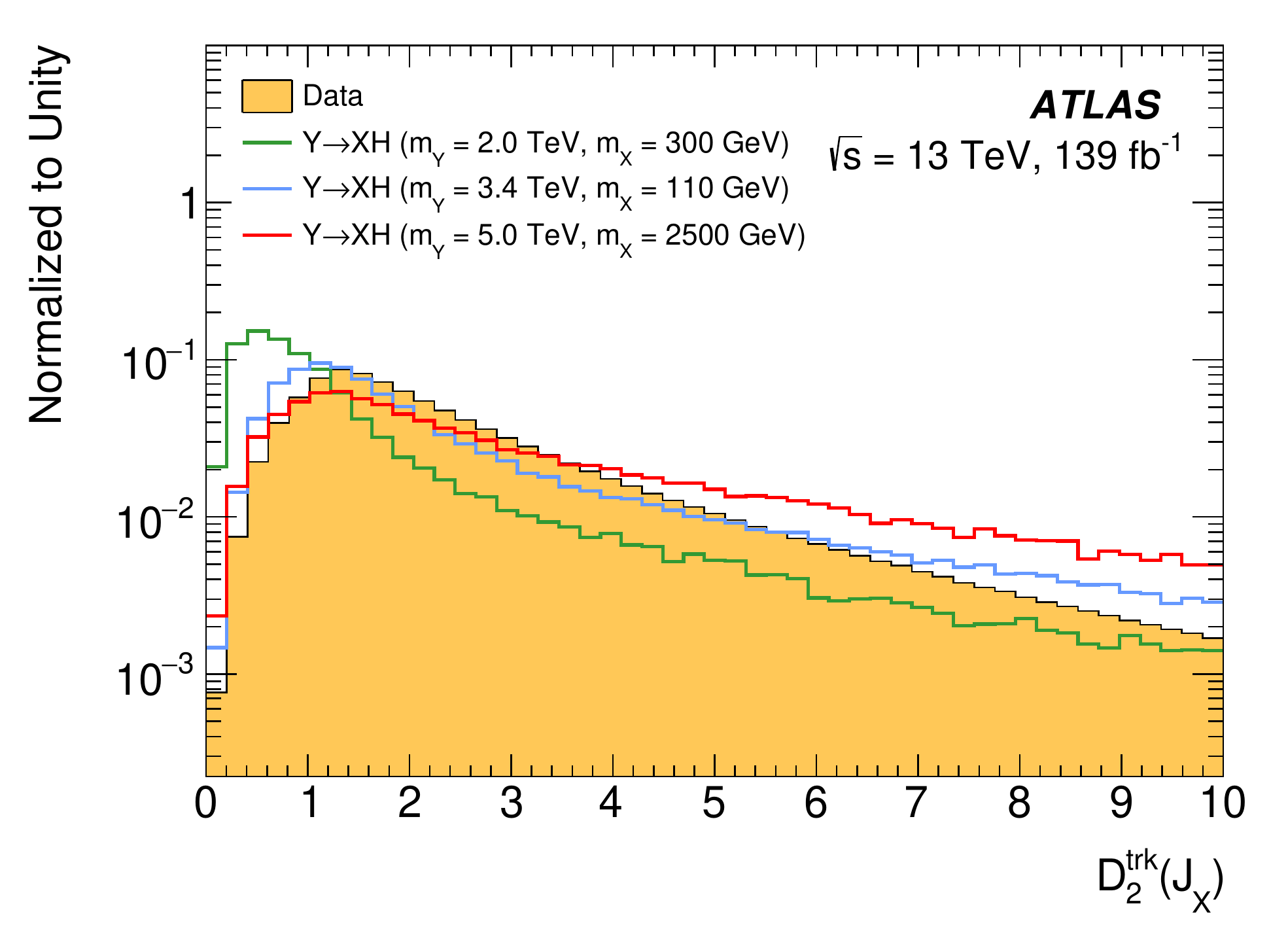}
\caption{\label{fig:X_vars} Distributions of $J_X$ candidate \dtrk in data after preselection requirements are applied. Also shown are three \yxh simulated signals, labeled by the masses of the $Y$ and $X$ particles. 
A value of $\dtrk = 1.2$ therefore separates the analysis events into merged ($< 1.2$) and resolved ($>1.2$) two-prong categories. All distributions are normalized to unity.}
\end{figure}
 
\subsection{$H$ selection}
 
The Higgs boson tagging is performed after the $J_X$ selection is applied, and sorts the events into the three analysis categories, for which the background is estimated. 
For all signal regions, a requirement $\D > 2.44$ providing a constant 60\% efficiency across jet \pt is applied to the Higgs boson candidate $J_H$, along with a mass window requirement of $75 < \mH < 145$~\GeV.
The data-driven background estimation uses events from the HSB of the Higgs boson candidate mass ($145 < \mH < 200~\GeV$), which are further separated into HSB0 and HSB1 depending on whether the Higgs boson candidate $J_H$ fails or passes the \D selection.
Validation is performed in the LSB, where the reconstructed Higgs boson mass must be between 65 and 75~\GeV.
LSB0 and LSB1 are similarly defined as having Higgs boson candidates that fail or pass the \D tagging criterion, respectively.
CR0 is defined as the set of events where the $H$ boson is in the SR mass window but fails the \D tagging. DNN-based reweighting is applied to CR0 to obtain the final background prediction in the SR. The contamination from signal events is found to be of the order of a few percent in the HSB, and no more than 10\% in the LSB, for all mass points in the \yxh grid.
 
In total, 3 SRs and 15 background estimation regions, 5 for each SR, are used in the analysis.
A summary of all region definitions is provided in Table~\ref{tab:selections}.
 
\begin{table*}[hbpt]
\centering
\caption{Preselection requirements, and optimized requirements defining the SRs and background estimation regions.}
\label{tab:selections}
\begin{tabular}{|c||c|c|c|c|c|}
\hline
Parameter & \multicolumn{5}{c|}{Preselection requirements} \\ \hline
\ \ \mjj \ [GeV] \ \ & \multicolumn{5}{c|}{ $> 1300$ }  \\
$p_\text{T}(J_1)$ \ [GeV] & \multicolumn{5}{c|}{ \phantom{0}$> 500$} \\
$m_J$ \ [GeV] & \multicolumn{5}{c|}{$m_{J_1} > 50$ || $m_{J_2} > 50$}  \\
\D & \multicolumn{5}{c|}{ \phantom{0\,}$> -2$}  \\
\hline \hline
& \multicolumn{5}{c|}{Signal regions} \\
&  \multicolumn{2}{c|}{Merged} & \multicolumn{2}{c|}{Resolved} &  \multicolumn{1}{c|}{ \ \ Anomaly \ \ } \\
\hline
\mH[GeV] & \multicolumn{5}{c|}{(75, 145) } \\
\D & \multicolumn{5}{c|}{$> 2.44$ } \\
\hline
\dtrk & \multicolumn{2}{c|}{$< 1.2$} & \multicolumn{2}{c|}{$> 1.2$} & \multicolumn{1}{c|}{ \ - \ } \\
$|\Delta y_{j_1,j_2}|$ & \multicolumn{2}{c|}{- } & \multicolumn{2}{c|}{ $< 2.5$ } & \multicolumn{1}{c|}{ \ - \ } \\
$p_\text{T}^{\text{bal}}$ & \multicolumn{2}{c|}{- } & \multicolumn{2}{c|}{ $< 0.8$ } & \multicolumn{1}{c|}{ \ - \ } \\
Anomaly Score (\SA) & \multicolumn{2}{c|}{-} & \multicolumn{2}{c|}{- } & \multicolumn{1}{c|}{ \ \ $> 0.5$ \ \ } \\
 
\hline \hline
& \multicolumn{5}{c|}{Background estimation regions} \\
&  CR0  & HSB0 & HSB1 & LSB0 & LSB1 \\
\hline
\mH[GeV] & \multicolumn{1}{c|}{ (75, 145) } & \multicolumn{2}{c|}{ (145, 200) } & \multicolumn{2}{c|}{ (65, 75) } \\
\hline
\D  & \multicolumn{1}{c|}{ $< 2.44$ } & \multicolumn{1}{c|}{$< 2.44$} &  \multicolumn{1}{c|}{$> 2.44$}& \multicolumn{1}{c|}{$< 2.44$} &  \multicolumn{1}{c|}{$> 2.44$} \\
\hline
\end{tabular}
\end{table*}

\subsection{Binning optimization}
 
The signal search in the two-dimensional space of \my vs.\ \mx applies a variable-width sliding window to the $J_X$ candidate mass spectrum, dividing the data into a series of overlapping \mx ranges in which the \mjj distribution is fitted.
The chosen binning for \my and \mx is based on the ATLAS detector's mass resolution for a generic narrow-width resonance, given by the fitted width of the Gaussian core of a double-sided Crystal Ball function.
Modifications are made to the binning to account for the limited number of data events.
In the \my spectrum, bins are widened at higher values of \my to ensure that at least one event is present in each bin when using LSB1 data.
For the \mx categories, an initial set of mass windows is generated from a linear fit of \mx resolution as a function of \mx.
The first window is chosen to have a bin center at $\mx = 65$~\GeV, the lowest generated signal $X$ mass, and its width is chosen to be twice the resolution obtained from the fit.
Subsequent windows, increasing in \mx, are chosen in the same way, where their bin centers are selected to be higher than the previous window's bin center by an amount that is equivalent to half of the previous window's mass resolution.
The widths of high-mass windows are expanded symmetrically, based on event counts in LSB1, until at least 10 events are present in each final \mx bin.
In order to avoid having duplicate windows, all resulting windows which share at least one bin edge are replaced by a single window encompassing their union.
Finally, to accommodate the highest signal mass point at $\mx = 3000$~\GeV, the last bin's upper edge is placed at 3200~\GeV.


\section{Background estimation}
\label{sec:bkgest}
 
The expected SM background that contributes in the SRs arises mainly from high-\pt multijet events, with jets that imitate the $b\bar{b}$ or $X$ topologies.
Simulation of such QCD processes is known to suffer from mismodeling and is computationally expensive.
Therefore, in this analysis the background estimation is fully data-driven and uses regions that are orthogonal to the SR because of their Higgs boson jet criteria.
 
The shape of the expected \mjj distribution in the SR is obtained from data in CR0, and weights are derived that can be applied to HSB0 to reproduce the shape found in HSB1.
This procedure is validated by applying the weights to data in LSB0 and comparing the resulting \mjj spectrum with that observed in LSB1 data.
The weights are generated inclusively in the \textit{X} candidate selection, and applied separately to create three background predictions, one for each SR.
 
The reweighting function is defined as the ratio of the multidimensional probability distribution function (pdf) of data in HSB1 to that of data in HSB0.
In this analysis, the statistical procedure of \textit{direct importance estimation}~\cite{hastie01statisticallearning, 10.5555/2181148} is utilized, where the ratio is estimated directly from data without having to analytically compute each individual PDF.
It is implemented via the training of a DNN, where a custom square-root loss function~\cite{Nachman_2021} is chosen to produce weights that can accurately reproduce the observed ratio in data.
The DNN outputs event-level weights that are assumed to be approximately independent of \mH, an assumption that can be verified using events in the LSB, and thus can be applied to an untagged region to produce the \mjj shape in the corresponding \Hbb-tagged region.
 
Events are considered for training the DNN if they pass the analysis preselection, satisfy $145 < \mH <175$~\GeV, and additionally have at least two track-jets associated with the Higgs boson candidate $J$.
They are modeled as a set of variables, namely the transverse momentum, pseudorapidity, azimuthal angle $\phi$ and energy of the Higgs boson candidate; the number of tracks associated with the Higgs boson candidate; and the transverse momentum, pseudorapidity, $\phi$ and mass of the first two track-jets associated with the Higgs boson candidate, ordered in \pT.
Each variable \textit{x}  is standardized with the transformation $x=(x-\mu )/\sigma$, where $\mu$ and $\sigma$ are the mean and the standard deviation of the distribution of the \textit{x}  variable.
 
The DNN is built using a fully connected sequential model from \textsc{Keras}~\cite{chollet2015keras} with 3 inner layers, each with 20 neurons and a rectified linear unit activation function.
In order to reduce the problem of overfitting during training, 10\% of connections among inner layers are randomly severed (called \enquote{dropout}).
The last layer has a single output with a simple linear activation function.
The model is trained using the Adam optimizer~\cite{adam} in \textsc{Keras}, with \textsc{Tensorflow}~\cite{tensorflow2015-whitepaper} as its backend.
Training is performed using a batch size equal to the full dataset size for 1600 epochs, with early stopping at 100 epochs if the value of the loss calculated on the validation dataset does not decrease for 100 subsequent epochs.
 
The output weights of the DNN are validated using data in the LSB.
Figure~\ref{fig:bkg} shows the impact of the reweighting on distributions of several key analysis variables, using the two-prong merged LSB VR as an example.
Three curves are shown for each variable, comparing the LSB0 data, before and after DNN reweighting is applied, with the target data distribution in LSB1.
These variables are chosen so as to focus on kinematic variables of the Higgs boson, whose properties are used to distinguish the CR from the SR, and variables involved in the final statistical treatment.
Good agreement between the reweighted shape and the true tagged data is observed for each variable, suggesting a robust background model.
Since the training is performed inclusively in the $X$-tagging, the same conclusion holds for the anomaly and two-prong resolved LSB regions, which is verified in comparisons of distributions between the reweighted LSB0 and LSB1 in those regions.
Also, since the reweighting is applied to \mjj distributions after the \textit{X}-tagging selection is applied, no extrapolation across \dtrk or \SA is required, and the method shows similar background modeling across all three SRs.
Any minor residual differences between predicted background and data in the SRs are covered by the nonclosure systematic uncertainty, described in Section~\ref{sec:syst}.
 
\begin{figure}[!htbp]
\centering
\includegraphics[width=0.45\textwidth]{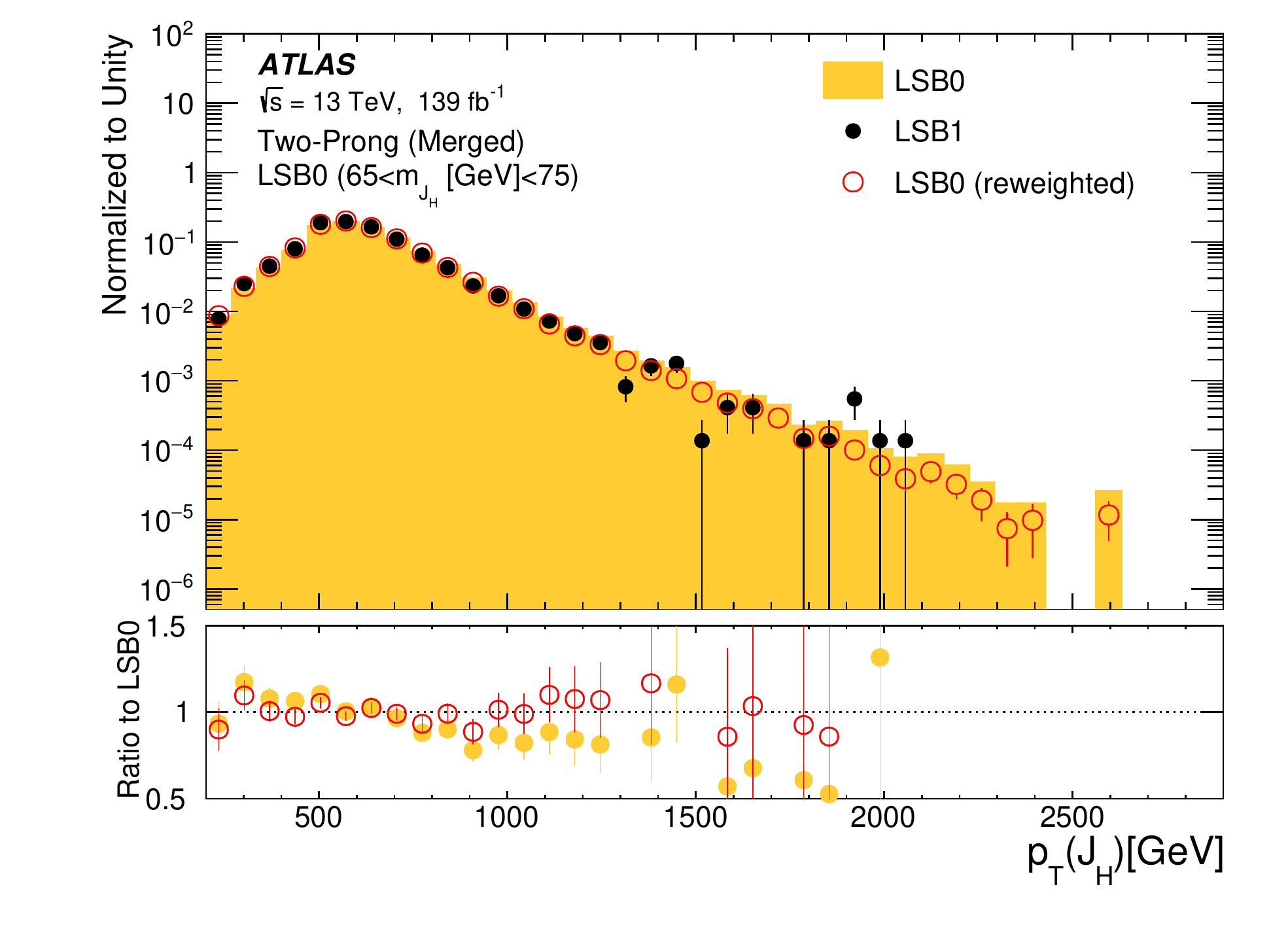}
\includegraphics[width=0.45\textwidth]{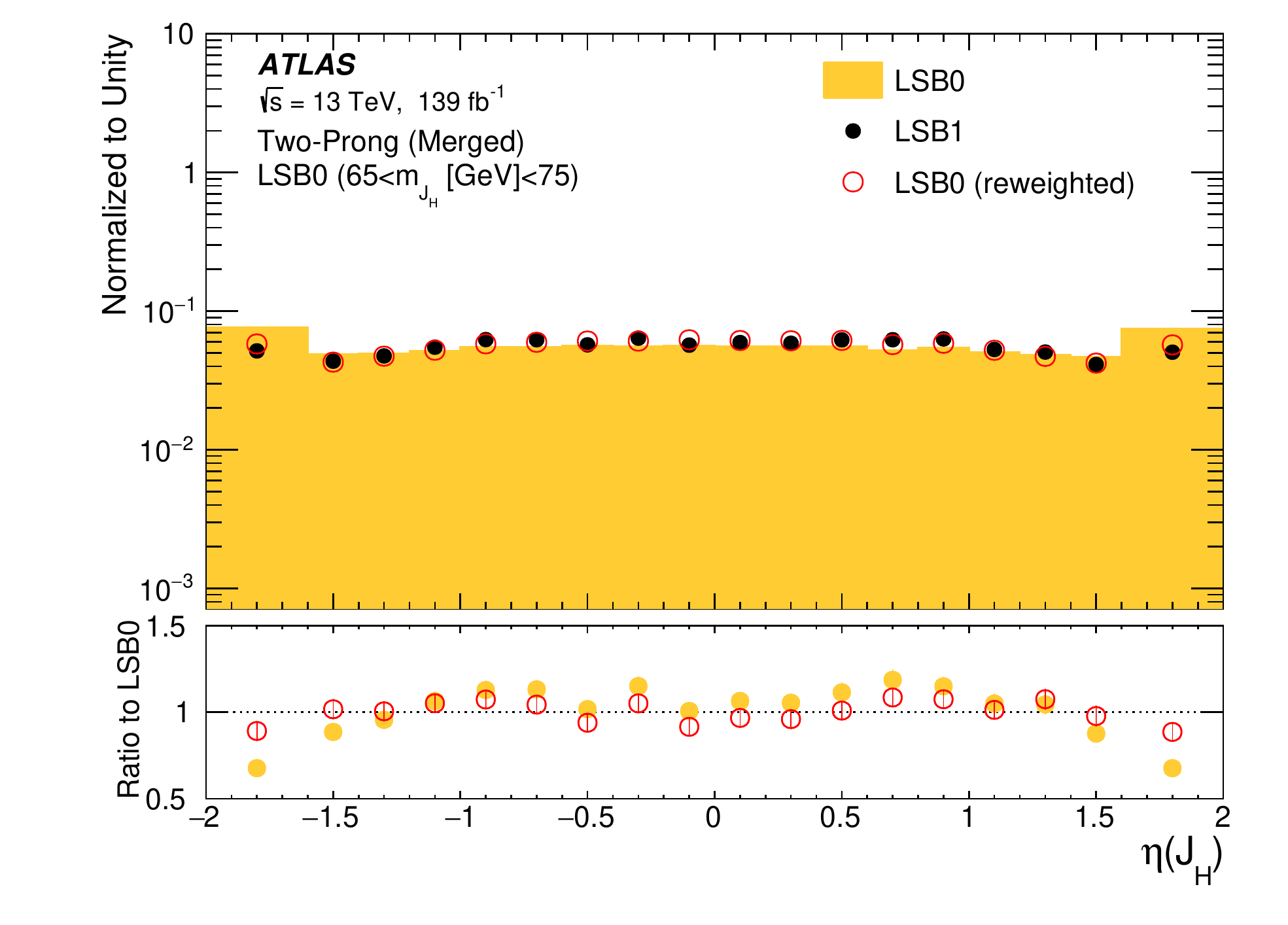}
\includegraphics[width=0.45\textwidth]{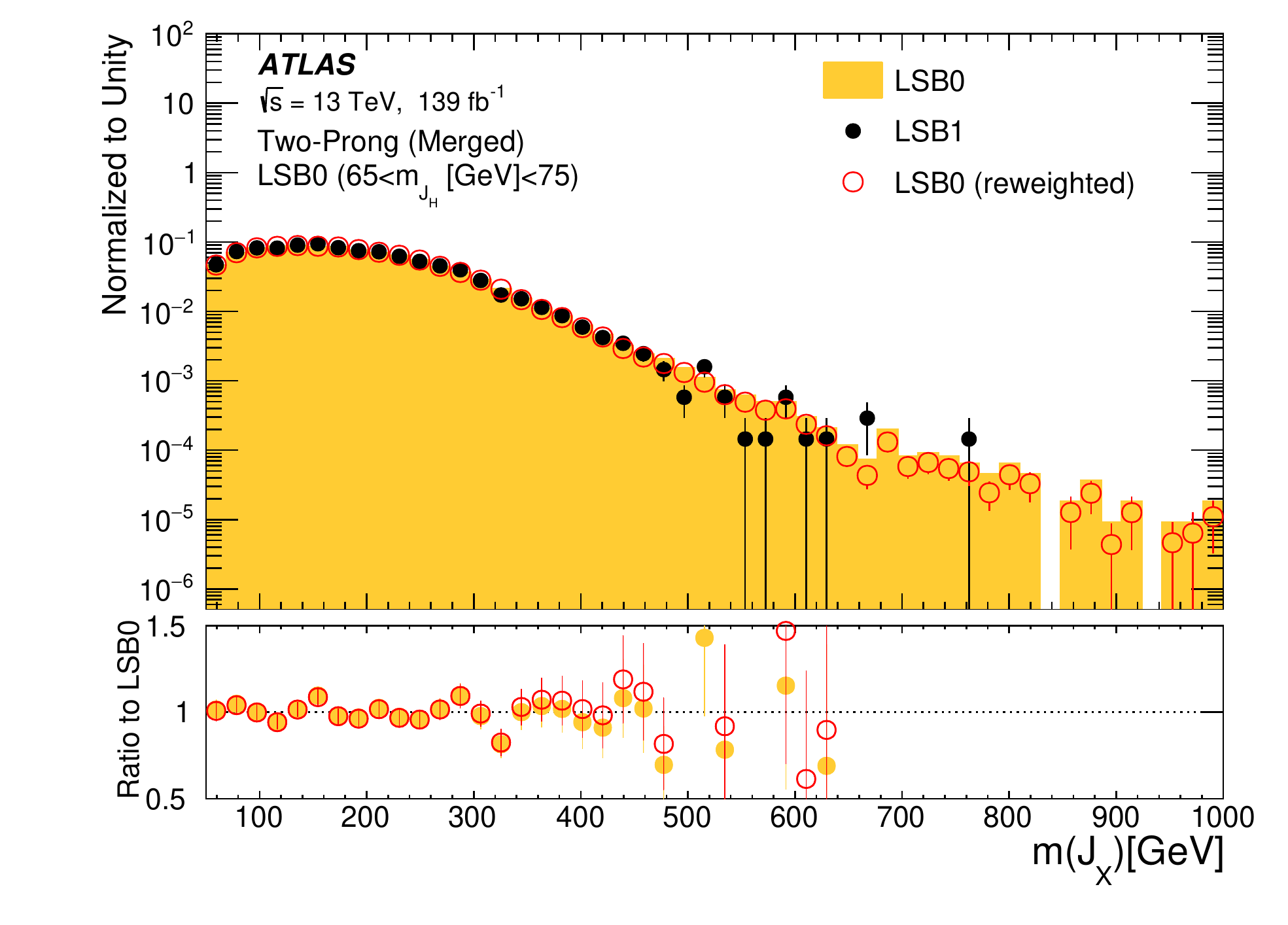}
\includegraphics[width=0.45\textwidth]{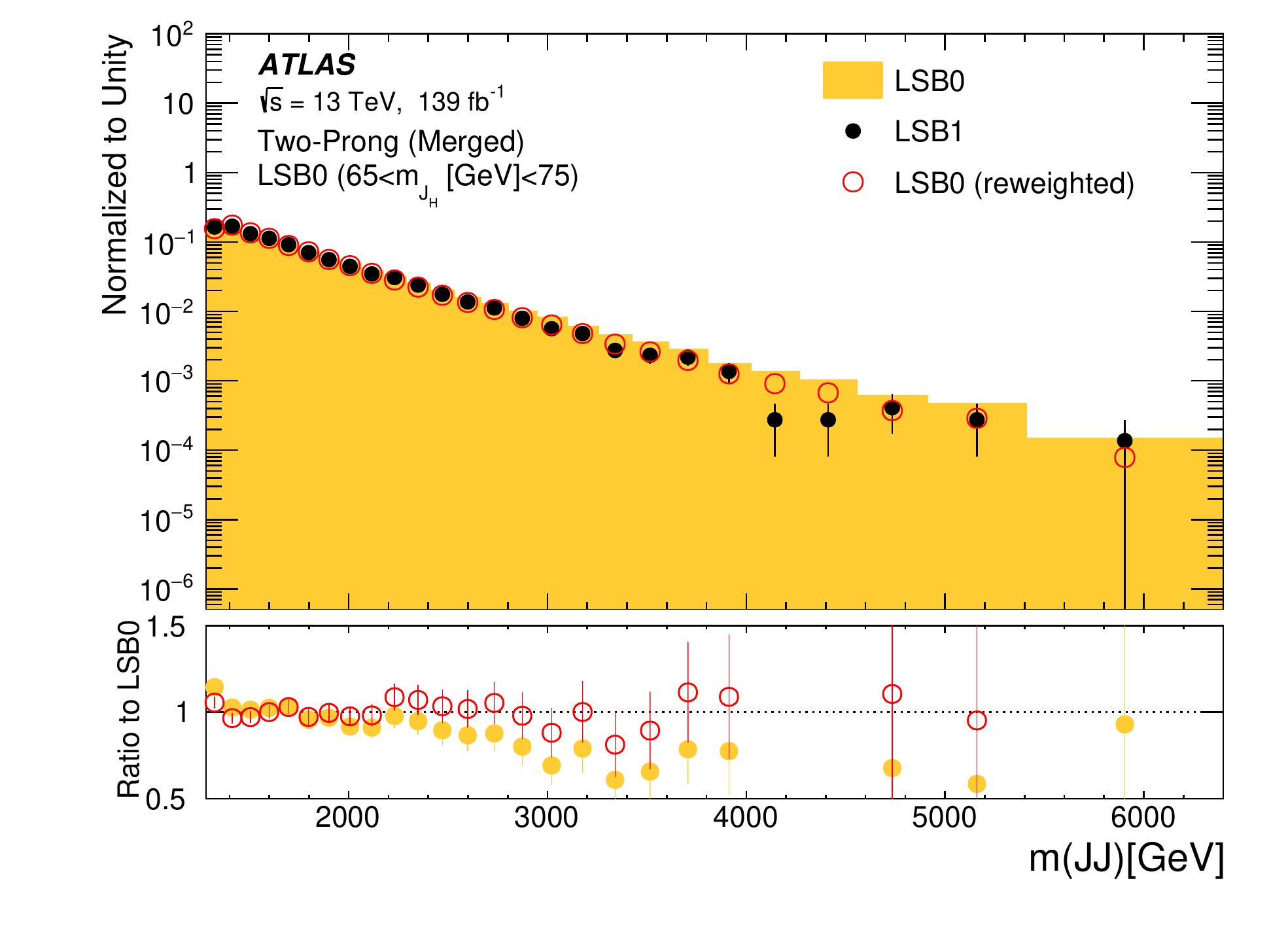}
\caption{Distribution of (a) \pt and (b) $\eta$ for the \textit{H} candidate $J$, along with (c) the mass of $J_X$ and (d) \mjj in the merged LSB validation region, overlaying data from LSB1 (solid black points) with the data in LSB0 shown before (orange histogram) and after (red circles) reweighting. The ratio of the LSB1 data to both the LSB0 data (orange) and the reweighted LSB0 data (red) is shown in the lower panel. Error bars indicate statistical uncertainties.
\label{fig:bkg}}
\end{figure}


\section{Systematic uncertainties}\label{sec:syst}
 
Systematic uncertainties affect both the data-driven background estimate and the simulated signal.
All background uncertainties are estimated using an \mjj shape that is inclusive in \mx, and applied to each \mx category fully correlated among \mjj bins.
Background shape variations are derived from three sources.
 
The first is from training the DNN in an alternative region of $165 < \mH < 200$~\GeV\ to account for the way in which phase-space uncertainties may affect the obtained weights.
This region has approximately the same number of events and tagging efficiency as the nominal training region, helping to isolate the way in which changing only the DNN model affects the weights.
Upward and downward variations are defined by symmetrizing the shape difference in \mjj between the two different models, yielding an effect of $\mathcal{O}(1{-}10)$\% across the distribution.
 
Another DNN variation accounts for the finite size of the training sample and the random initialization of the weights.
It is estimated with a bootstrap procedure~\cite{efron, michelucci2021estimating} where a set of 100 bootstrap networks are trained. For each training, the training dataset is varied by resampling it with replacement, assigning to each event a new weight that comes from a Poisson variable with a mean of one. Two additional templates are formed to create a symmetric error band by taking the median weight for each event and adding or subtracting half of the interquartile range.
This corresponds to a $\mathcal{O}(1)$\% effect across \my.
 
Lastly, a nonclosure uncertainty is included to cover modeling discrepancies that may arise when weights derived from the NN training in the HSB are extrapolated to the LSB, and subsequently to the SR.
It is defined by the symmetrized shape difference between the data and predicted background in the LSB.
To avoid being sensitive to statistical fluctuations in the LSB, smoothing is applied to the variation, where it is rebinned to reduce the relative statistical uncertainty.
The nonclosure uncertainty is negligible for low \mjj and rises to $\mathcal{O}(10)$\% in the \mjj tails.
 
Normalization and shape uncertainties are applied to the simulated signals.
An uncertainty of 1.7\% is applied to the luminosity measured with the LUCID-2 detector~\cite{lumi}.
The uncertainty from the trigger selection is negligible, and thus not included, since the requirement of $\mjj >1.3$~\TeV\ ensures that the trigger is fully efficient.
 
Scale factors (SFs) computed to match the \Hbb tagger efficiency between data and simulation are applied to the signal template, and their uncertainties are similarly propagated to the signal normalization.
These SFs are computed using the methodology described in Ref.~\cite{xbb_calib}, with the substitution of an updated \D that includes the $\eta$-reweighting of inputs as described in Section~\ref{sec:selection}.
They are binned in large-$R$-jet \pt, where the SF for the highest \pt bin is extrapolated to cover the upper end of the \pt region probed by the analysis selection.
Agreement between data and MC simulation for these SFs is verified with $Z\rightarrow b\bar{b}$ events in calibration studies.
The SFs are bounded by approximately 1.1 and 1.4 across all \pt bins, with uncertainties in the range of approximately $\pm$(0.3--0.5).
 
Instrumental systematic uncertainties arise from uncertainty in the jet \pt scale and resolution, for both the large-$R$ and small-$R$ jets, and affect the shape of the \mjj distribution.
Uncertainty in the large-$R$-jet \pt scale is an important effect in the search for resonant structures in the presence of rapidly falling background spectra, as it can shift the peak of the resonance.
It is evaluated using tracking-to-calorimetric \pT double ratios between data and simulation, where any observed differences are assigned as baseline systematic uncertainties~\cite{PERF-2012-02}.
Even though the analysis relies on jets built from TCCs, the total \pt of the jet is still solely derived from calorimeter information, keeping it independent of the track-based jet \pt.
Past analyses have studied the possible impact of calorimeter vs.\ track-based \pt by cross-calibrating between per-jet TCC and calorimeter \pt and found it to be negligible~\cite{vvjj_fullrun2}.
The total jet-\pt-scale uncertainty varies with \pt and $\eta$ and is typically around $\pm$5\%.
Uncertainties from the reconstruction and modeling of tracks are also taken into account; these cover track reconstruction efficiency, impact parameter resolution, tracking in dense environments, fake-track rates, and sagitta biases.
The impact of the large-$R$-jet \pt resolution uncertainty is evaluated event-by-event by rerunning the analysis with an additional absolute 2\% Gaussian smearing applied to the input jets' \pt to degrade the nominal resolution.
Small-$R$-jet \pt scale and resolution uncertainties are similarly estimated by comparing data with simulation after applying in~situ corrections~\cite{JETM-2018-05}.
 
Several sources of theoretical uncertainty affecting the signal models were considered.
Uncertainties in the matrix element calculation are evaluated by varying the strong coupling constant (\alphas), the renormalization scale (\muR), and the factorization scale (\muF). The effects of uncertainties in the parton distribution functions
are evaluated by comparing the \mjj distributions obtained when using various alternative PDF sets and taking an envelope of these distributions, as prescribed by the PDF4LHC group~\cite{Butterworth:2015oua}. Generator-level variations of the A14 tune's parameter values are used to cover the initial- and final-state radiation (ISR and FSR) and multiple parton interaction (MPI) uncertainties.
An overall conservative 3$\%$ normalization uncertainty is applied to all signals as a result of ISR/FSR/MPI modeling effects.


\section{Statistical analysis}
\label{sec:statistical}
 
The statistical framework in the analysis is used to perform hypothesis tests in the SRs, testing the compatibility of the data with both the background-only and signal-plus-background hypotheses. The observable that is fitted is the \my distribution of the data in the SR.
This fit is repeated several times in overlapping bins of the \textit{X} candidate mass.
 
The parameter of interest in the statistical analysis is the signal strength $\mu$, defined as a scale factor multiplying the nominal yield predicted by a 1~pb signal cross section so as to match the observed number of signal events.
The background-only hypothesis corresponds to $\mu$ = 0. 
The normalization of the data-driven background estimate is allowed to float, with each normalization factor being fitted independently because the \mx categories overlap.
A test statistic based on the profile likelihood ratio calculated in the lowest-order asymptotic approximation is used to test the models corresponding to the signal grid.
Systematic uncertainties are included in the fit as nuisance parameters (NPs) with Gaussian constraints.
Both the signal strength and all signal systematic uncertainty NPs are correlated across the merged and resolved regions.
The significance of an observed excess of data over the background prediction is quantified by the local $p$-value, which is the probability that the background-only model produces an excess at least as large as the one observed.
 
\textsc{BumpHunter}~\cite{bumphunter} is used to find excesses in both the anomaly and two-prong SRs, with an algorithm that incorporates only the statistical uncertainty of the data and does not depend on a specific signal shape.
It outputs a $p$-value that provides a goodness-of-fit metric, along with an interval of the invariant mass corresponding to the largest deviation of the data from the background prediction.
It is thus used in all signal regions to search for significant deviations of the data from the background \my distribution.
Dedicated studies indicate the fits are dominated by statistical uncertainties, further motivating the discounting of systematic uncertainties in the \textsc{BumpHunter} fits.
In the anomaly SR, no subsequent fits are performed using the \yxh signal model, as this analysis is designed to keep signal-model dependence minimal and the two-prong regions are expected to have higher signal sensitivity across the \yxh grid.
However, given the absence of a significant excess in the two-prong regions, signal-plus-background fits are performed for each \yxh signal model, and 95\% confidence level (CL) upper limits on the signal cross section are set using a modified frequentist method (CL\textsubscript{s})~\cite{Read:2002hq}.

The background estimation and statistical treatment are validated by comparison with data in the signal-depleted LSB VR for each of the three analysis categories.
\textsc{BumpHunter} $p$-values from all \mx-bin fits in the VR of the anomaly region approximate a flat distribution between 0 and 1, indicating good background modeling with no systematic biases across the phase space.
Figure~\ref{fig:vr_excl} shows the post-fit background prediction compared with the data passing the anomaly, merged, and resolved selections, in an example \mx window between 284.5 and 322.5~\GeV\ where the whole kinematic phase space is well-populated.
Good modeling is observed, and no significant pulls or constraints are observed in the background NPs.
In the two-prong regions, checks are made for spurious signals in the LSB through a signal-plus-background fit using each generated signal model.
These checks indicate that no significant spurious signal is present beyond those likely to be produced by statistical fluctuations, with no systematic trend across the expected phase space.
 
\begin{figure}[hbpt]
\centering
\includegraphics[width=0.45\textwidth]{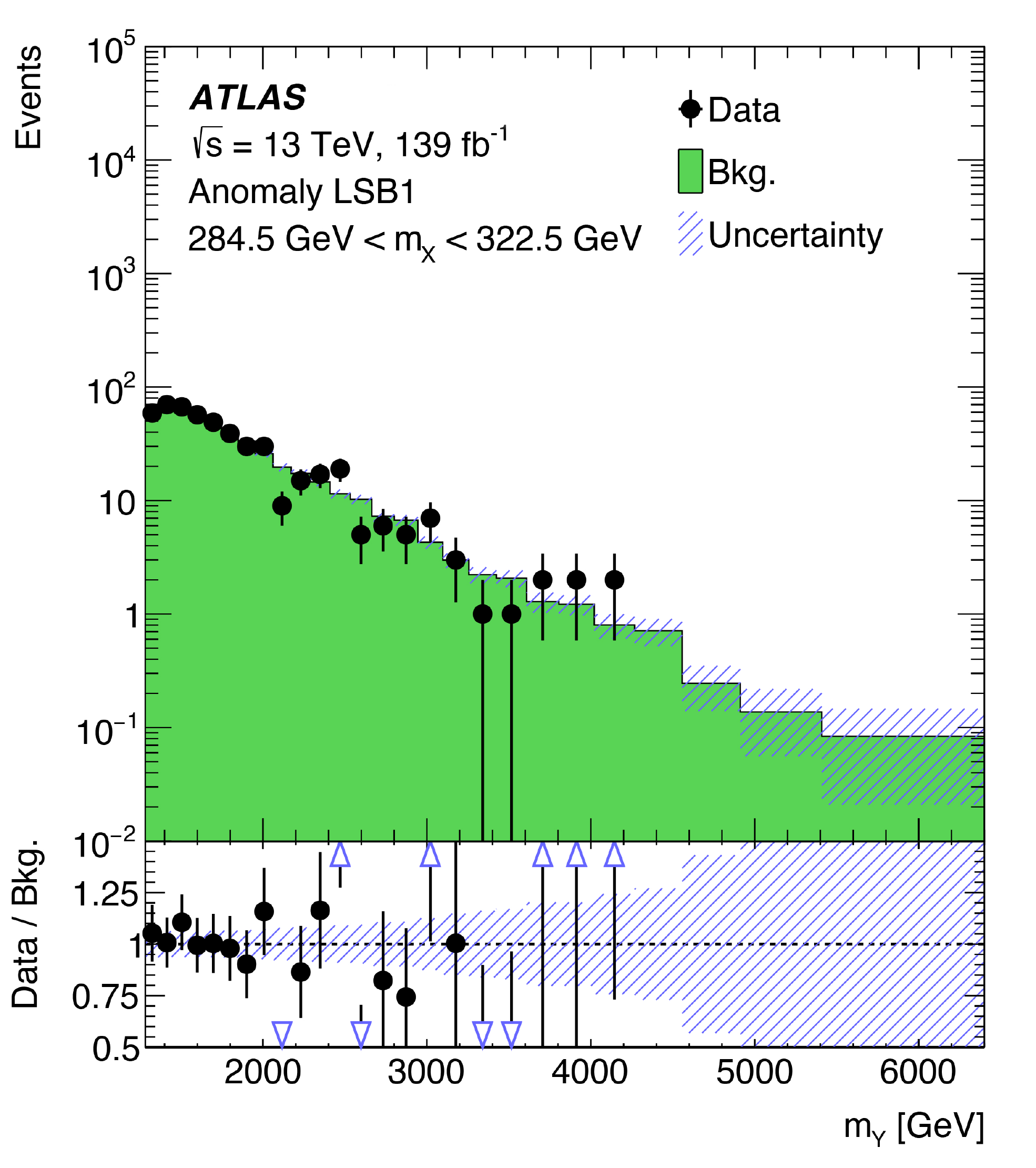}
\includegraphics[width=0.45\textwidth]{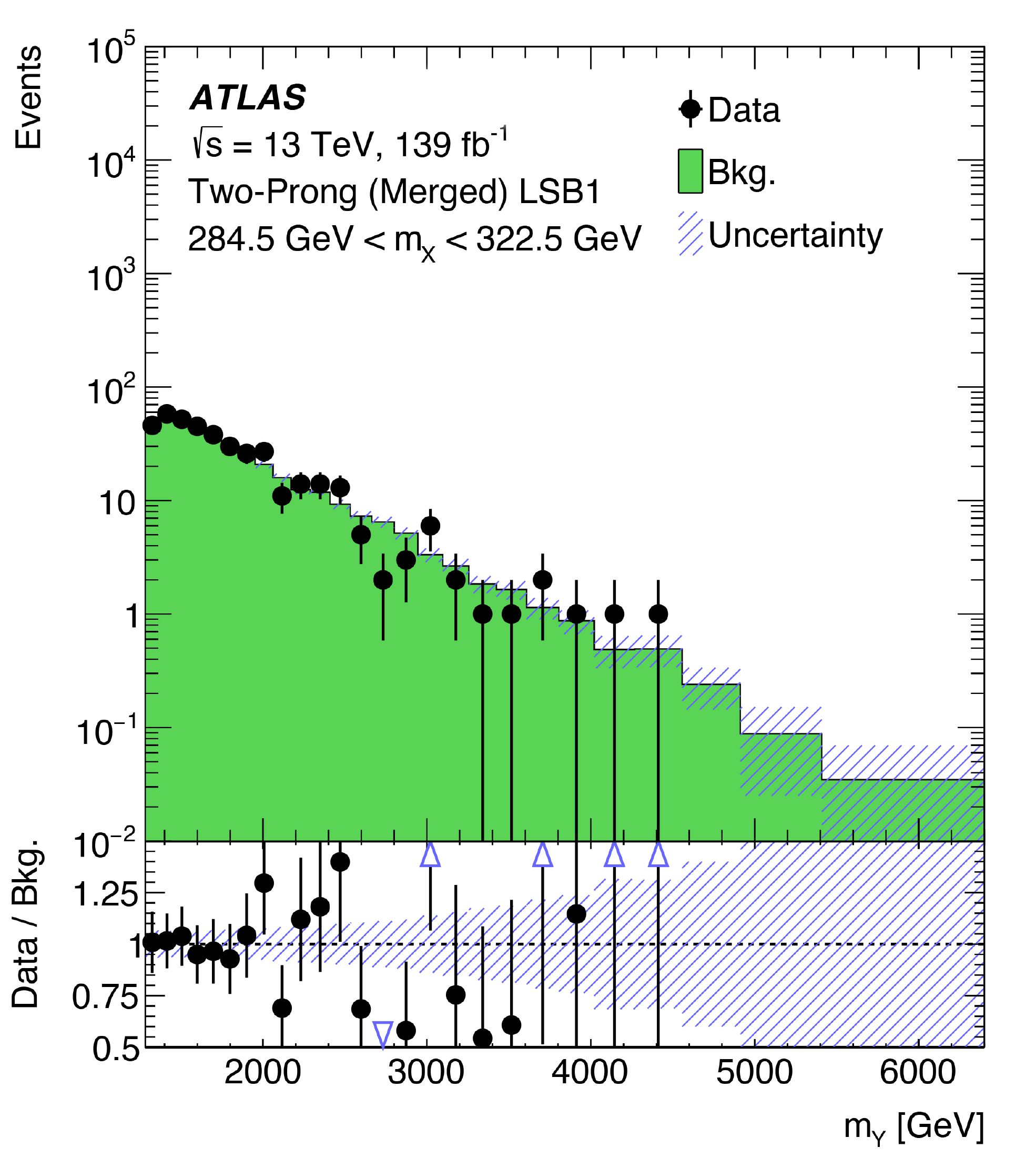}
\includegraphics[width=0.45\textwidth]{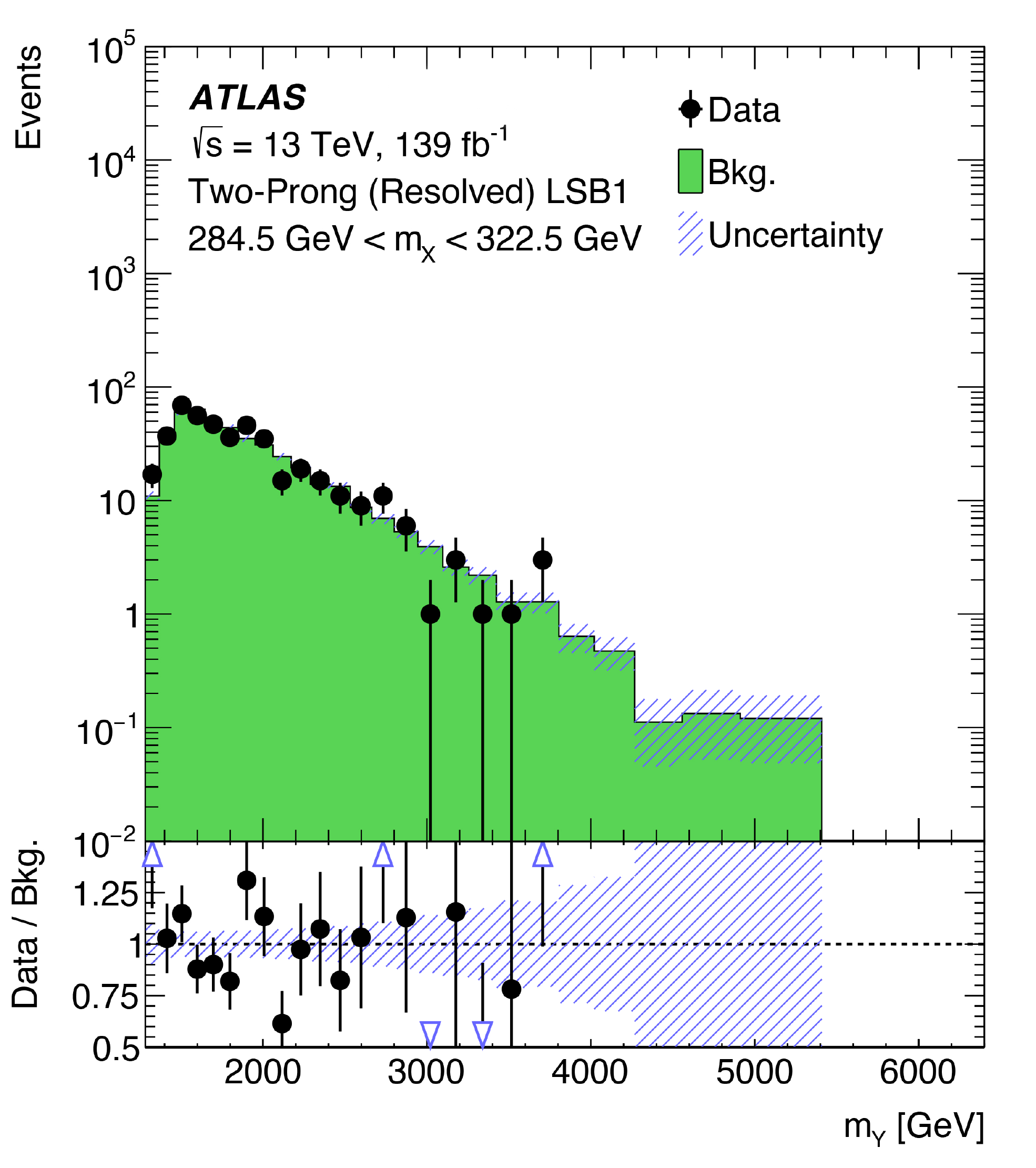}
\caption{\label{fig:vr_excl} Reconstructed \my~distributions of the background, as determined by a background-only fit, and data in the LSB1 VR, for the anomaly, two-prong merged, and two-prong resolved selections, in the \mx bin [284.5, 322.5]~\GeV. The ratio of the observed data to the background is shown in the lower panel. The uncertainty band includes both the statistical and systematic effects.}
\end{figure}


\section{Results}
\label{sec:result}
 
Results of background-only fits of the \my distribution across all \mx categories in the anomaly SR show the data to be compatible with the  expected backgrounds given their uncertainties.
Figure~\ref{fig:results_p0_scan} shows the distribution of $p$-values across \my and \mx bins, where the background is the result of a background-only fit performed with all statistical and systematic uncertainties, and all uncertainties are included in the $p$-value calculation.
The distribution of observed $p$-values is statistically compatible with the expectation, with a $\chi$-squared value per degree of freedom of 1.32 between the distributions.
 
\begin{figure}[hbpt]
\centering
\includegraphics[width=0.95\textwidth]{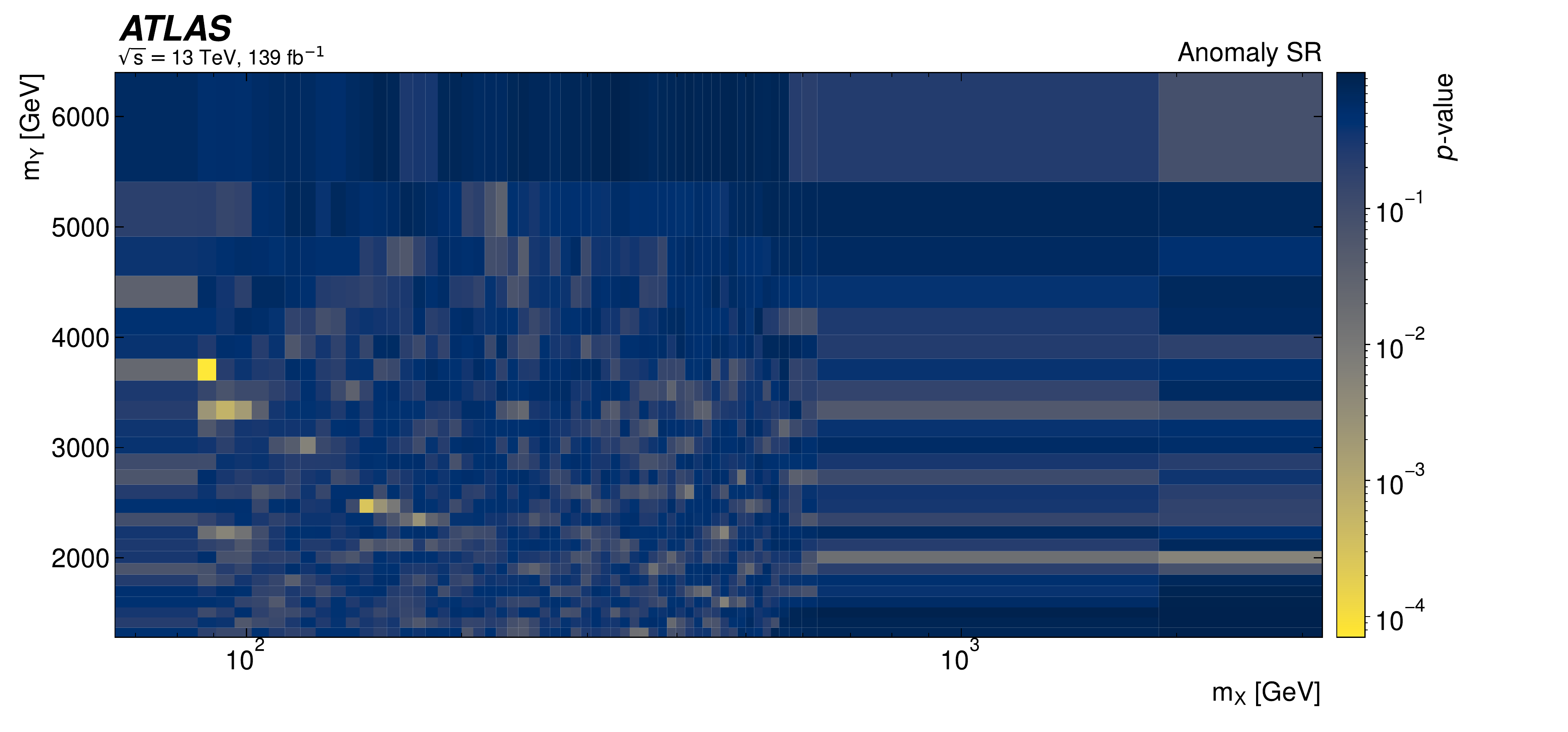}
\caption{\label{fig:results_p0_scan} The distribution of observed $p$-values across all \my and \mx bins in the anomaly signal region, comparing data with the background estimates generated by a background-only fit, displayed in the two-dimensional (\mx, \my) grid. The $p$-value calculation is performed at the center of each \mx~bin, and all statistical and background systematic uncertainties are considered. The lowest observed $p$-value corresponds to the bin with \my within [3608, 3805]~\GeV\ and \mx within [75.5, 95.5]~\GeV.}
\end{figure}
 
The largest observed excess in the anomaly SR is in the \mx window [75.5, 95.5]~\GeV.
In this window, the greatest single-bin excess is found for \my between 3608 and 3805~\GeV.
The \textsc{BumpHunter} interval with the greatest significance covers the \my range between bin edges of 3424 and 3805~\GeV.
The \my distribution corresponding to this \mx window is shown in Figure~\ref{fig:as_excess_mjj}, along with the post-fit expected background.
Studies of the individual jet mass distributions in this \mx category do not reveal any excesses in data that are consistent with a resonant particle decay.
Given the number of individual search regions in this analysis, the impact of the trials factor is significant.
A calculation is made to determine the global significance of the greatest single-bin deviation, accounting for all the overlapping \mx bins. First, the overlapping bin edges are used to define exclusive, nonoverlapping bins in \mx, and an integer is drawn from a Poisson distribution with a mean equal to the background expectation in each exclusive (\my, \mx) bin. Next, this yield is summed across exclusive bins to create a toy estimate for each overlapping bin in which the $p$-value is computed. Finally, this procedure is repeated $N$ times, where $N$ is the number of events, inclusively across all exclusive, nonoverlapping bins.
This calculation yields a global significance of 1.43$\sigma$ for the excess in the [3608, 3805]~\GeV bin of \my.
 
\begin{figure}[hbpt]
\centering
\includegraphics[width=0.5\textwidth]{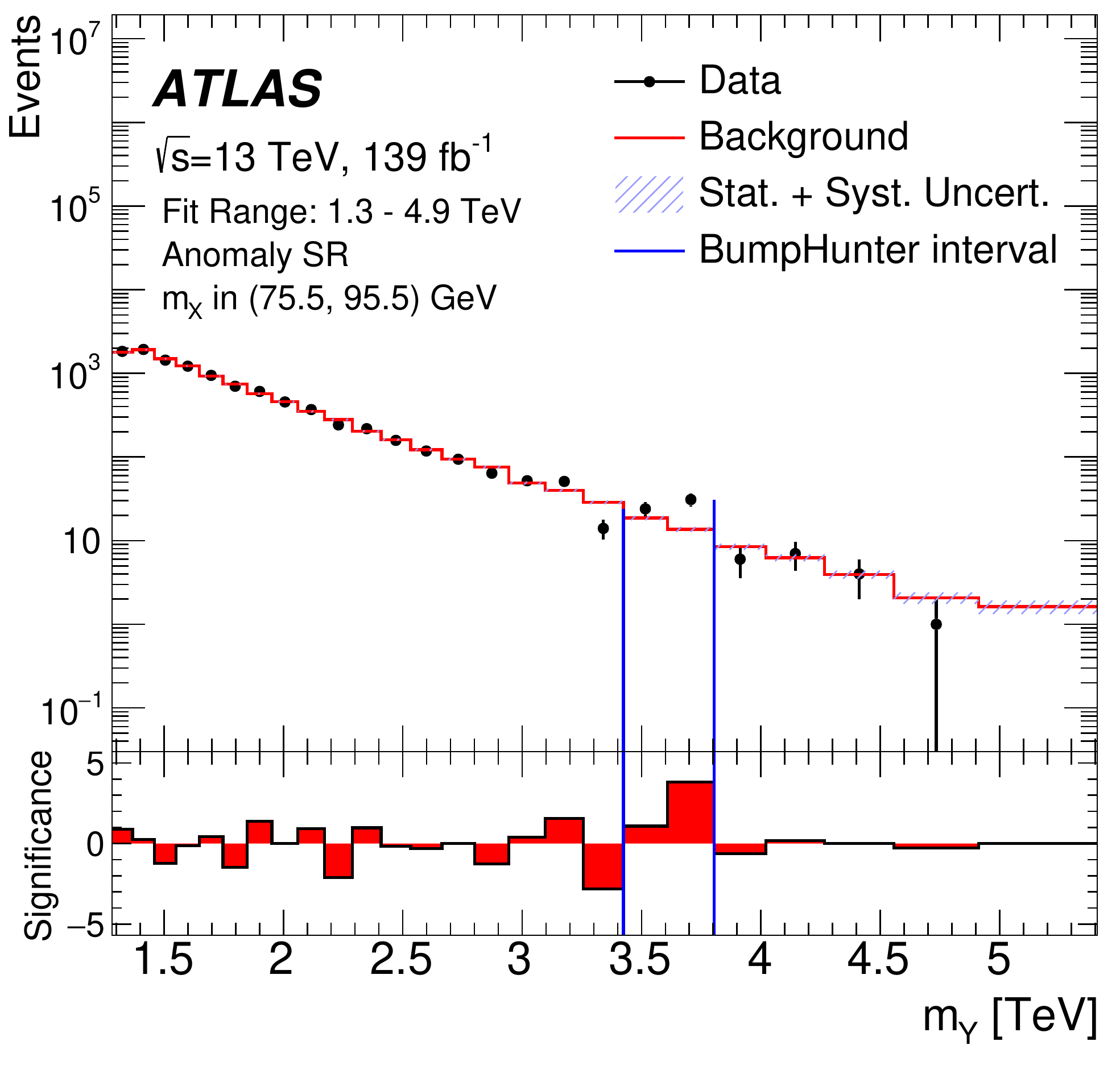}
\caption{\label{fig:as_excess_mjj} The \my distribution associated with the \mx bin [75.5, 95.5]~\GeV\ in the anomaly SR, where the background is determined by a background-only fit to data with both the statistical and systematic uncertainties included.
The $p$-value and the lower panel of per-bin significances are computed with statistical uncertainties only. The single \my bin with the most significant excess is [3608, 3805]~\GeV. Incorporating the trials factor for all \my and overlapping \mx bins in the search, this excess has a global significance of 1.43$\sigma$.}
\end{figure}
 
Results for the two-prong SRs are similarly derived by performing background-only fits and scanning with \textsc{BumpHunter} for incompatibility with data.
No significant deviations of data from the predicted background are observed beyond expected statistical fluctuations, in either the merged or resolved SR.
An example \my distribution in both the merged and resolved SRs for the \mx bin [284.5, 322.5]~\GeV\ is shown in Figure~\ref{fig:sr_excl}, along with the background estimate that is determined from a background-only fit accounting for all uncertainties.
\begin{figure}[hbpt]
\centering
\includegraphics[width=0.45\textwidth]{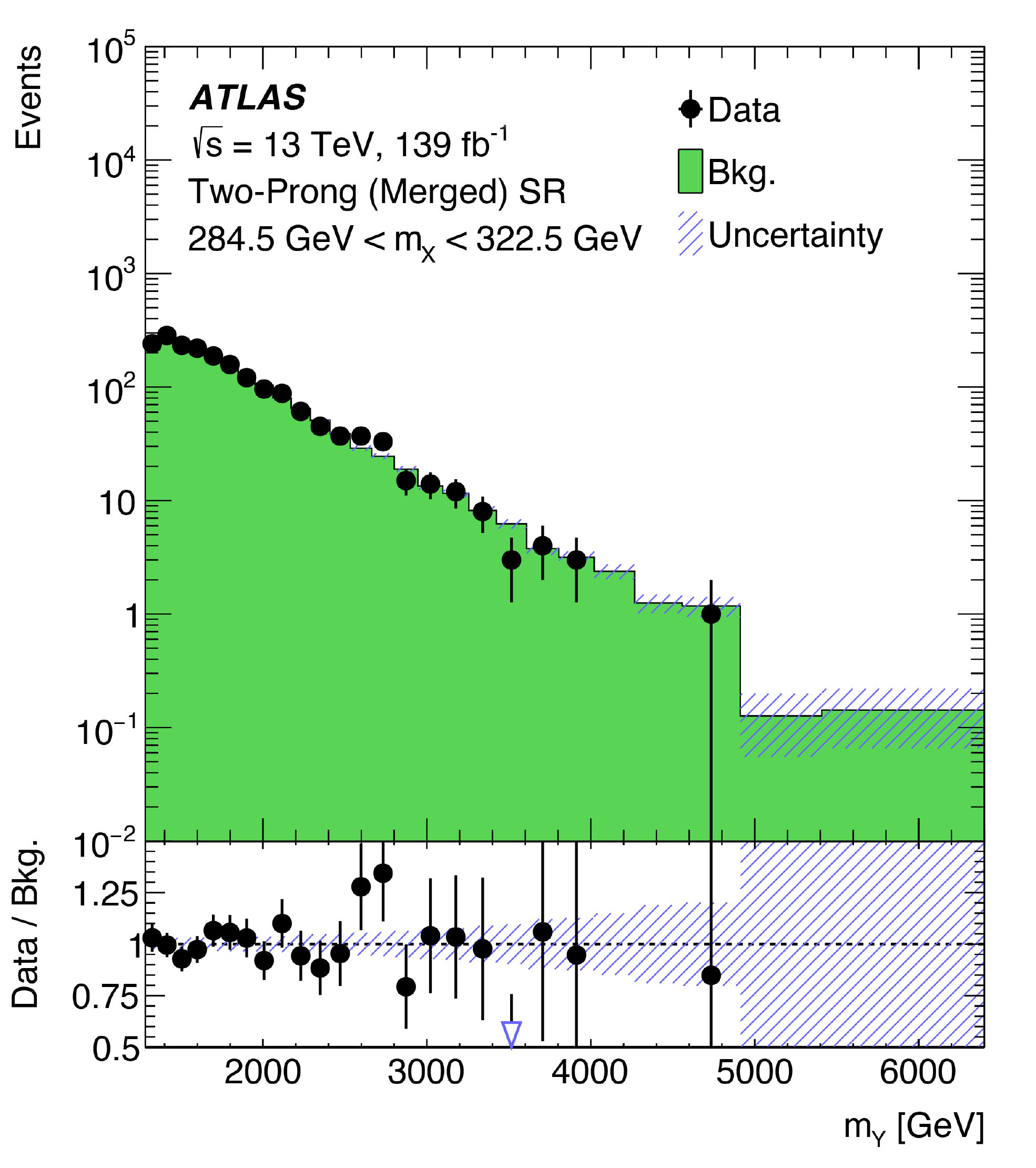}
\includegraphics[width=0.45\textwidth]{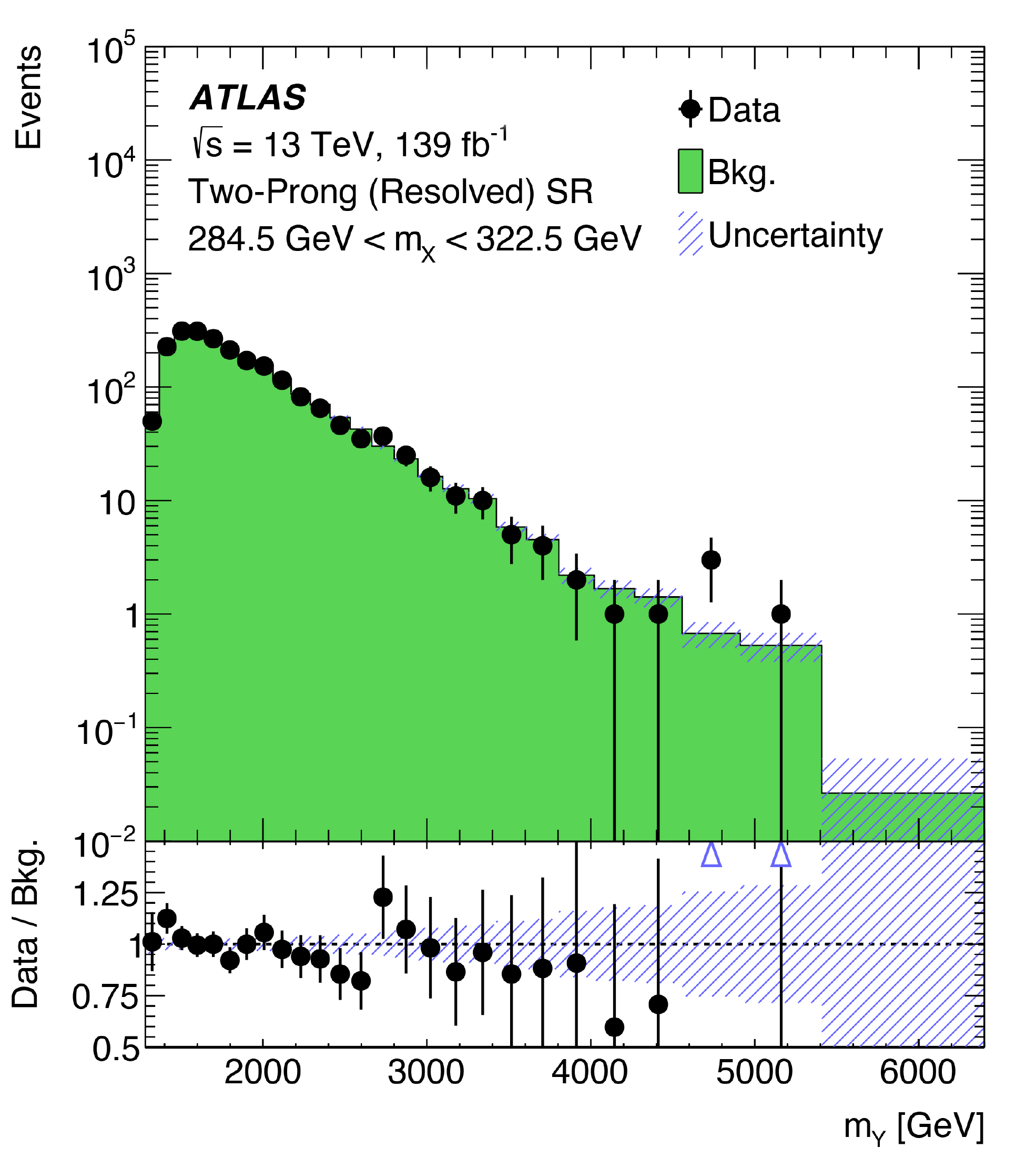}
\caption{\label{fig:sr_excl} Reconstructed \my distributions of background and data in the SR, for the merged (left) and resolved (right) selections, in the \mx bin [284.5, 322.5]~\GeV. The background is determined by a background-only fit to the data with all statistical and systematic uncertainties included. The ratio of the observed data to the background is shown in the lower panel. The uncertainty band includes both the statistical and systematic effects.}
\end{figure}
Figure~\ref{fig:results_excl} shows a summary of the per-bin $p$-values in each \my bin for selected \mx bins that are centered on key \textit{X} candidate mass hypotheses, namely at the $W$, $Z$, and \textit{H} boson masses.
 
\begin{figure}[hbpt]
\centering
\includegraphics[width=0.45\textwidth]{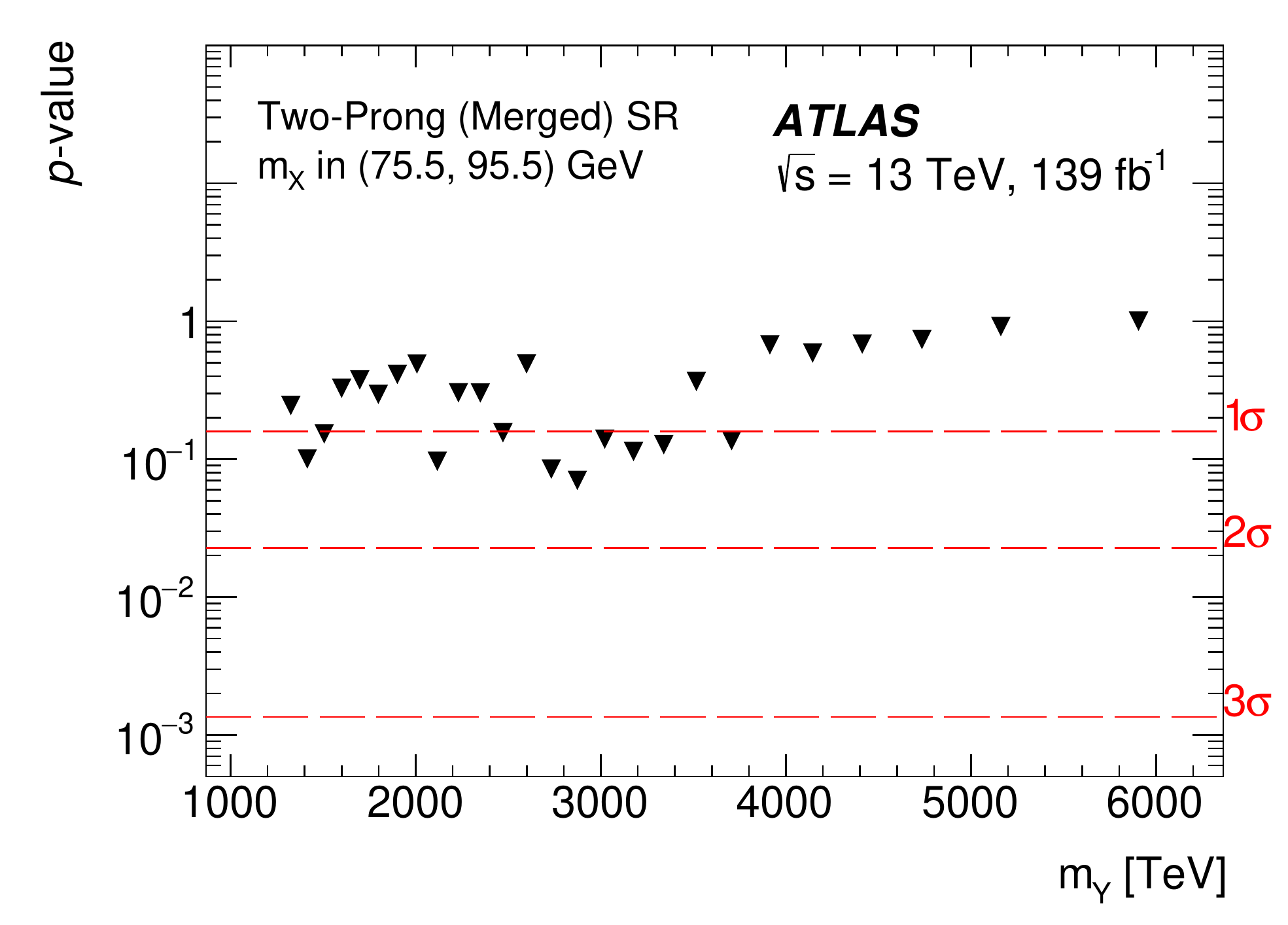}
\includegraphics[width=0.45\textwidth]{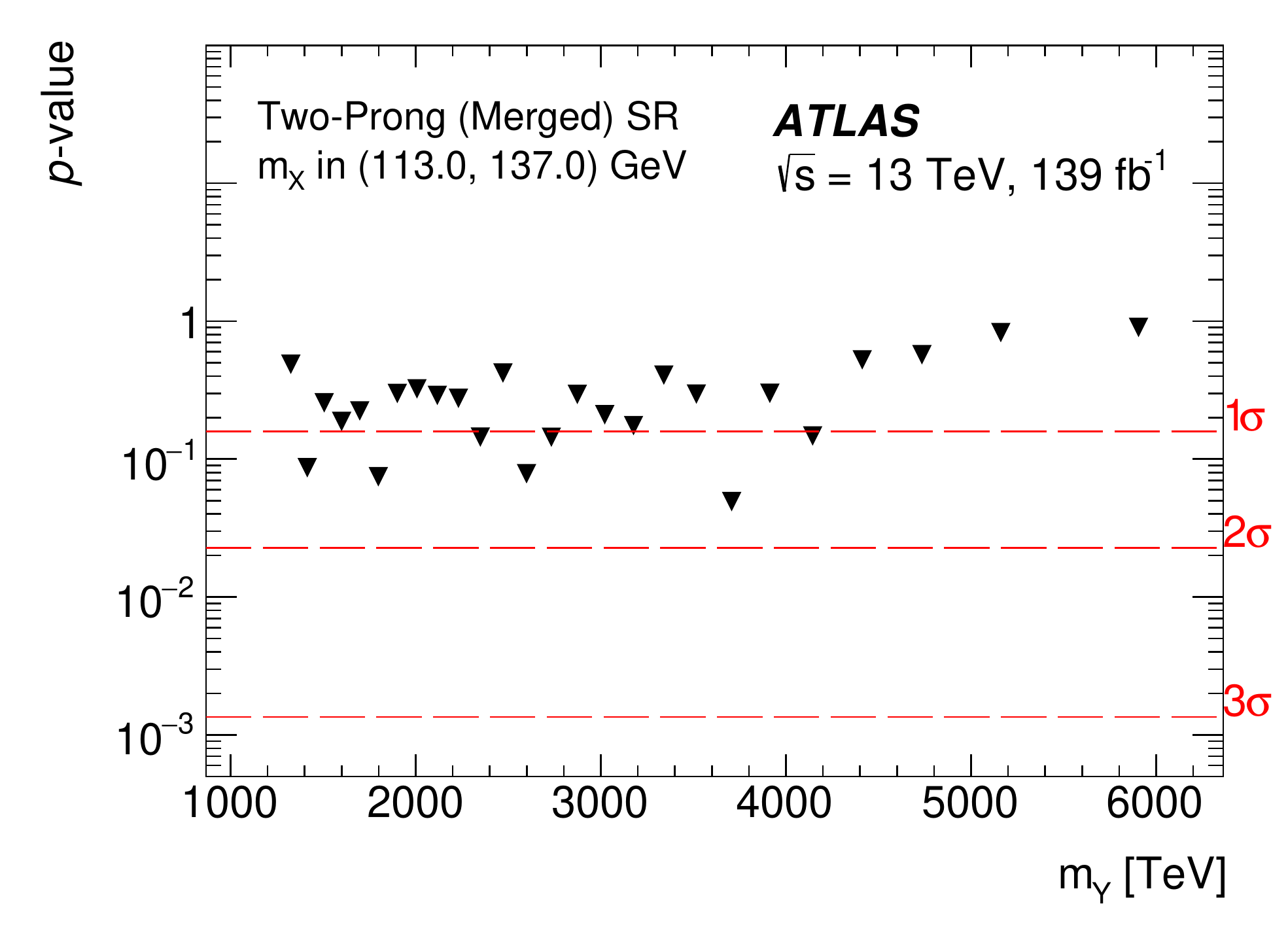}
\includegraphics[width=0.45\textwidth]{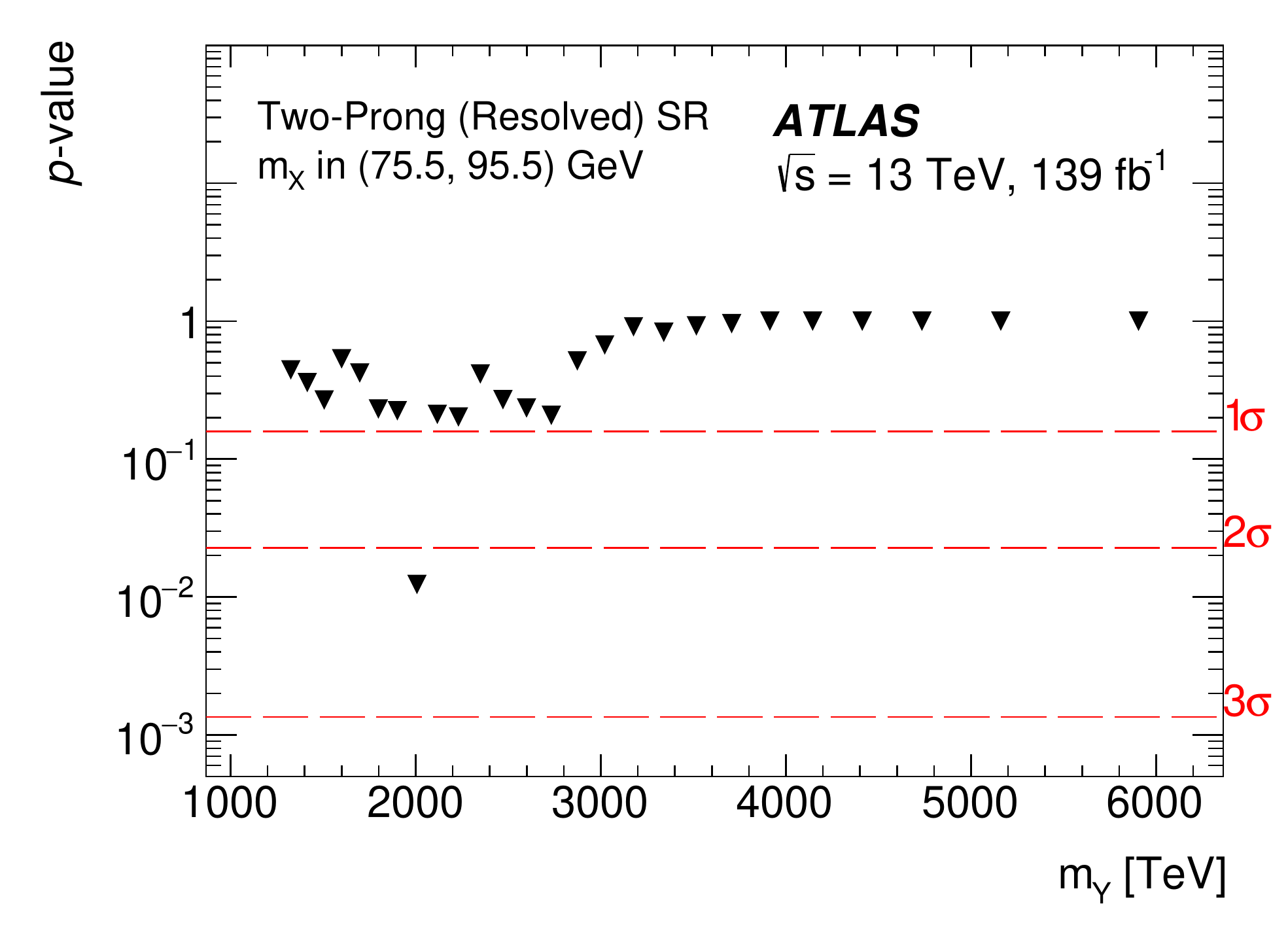}
\includegraphics[width=0.45\textwidth]{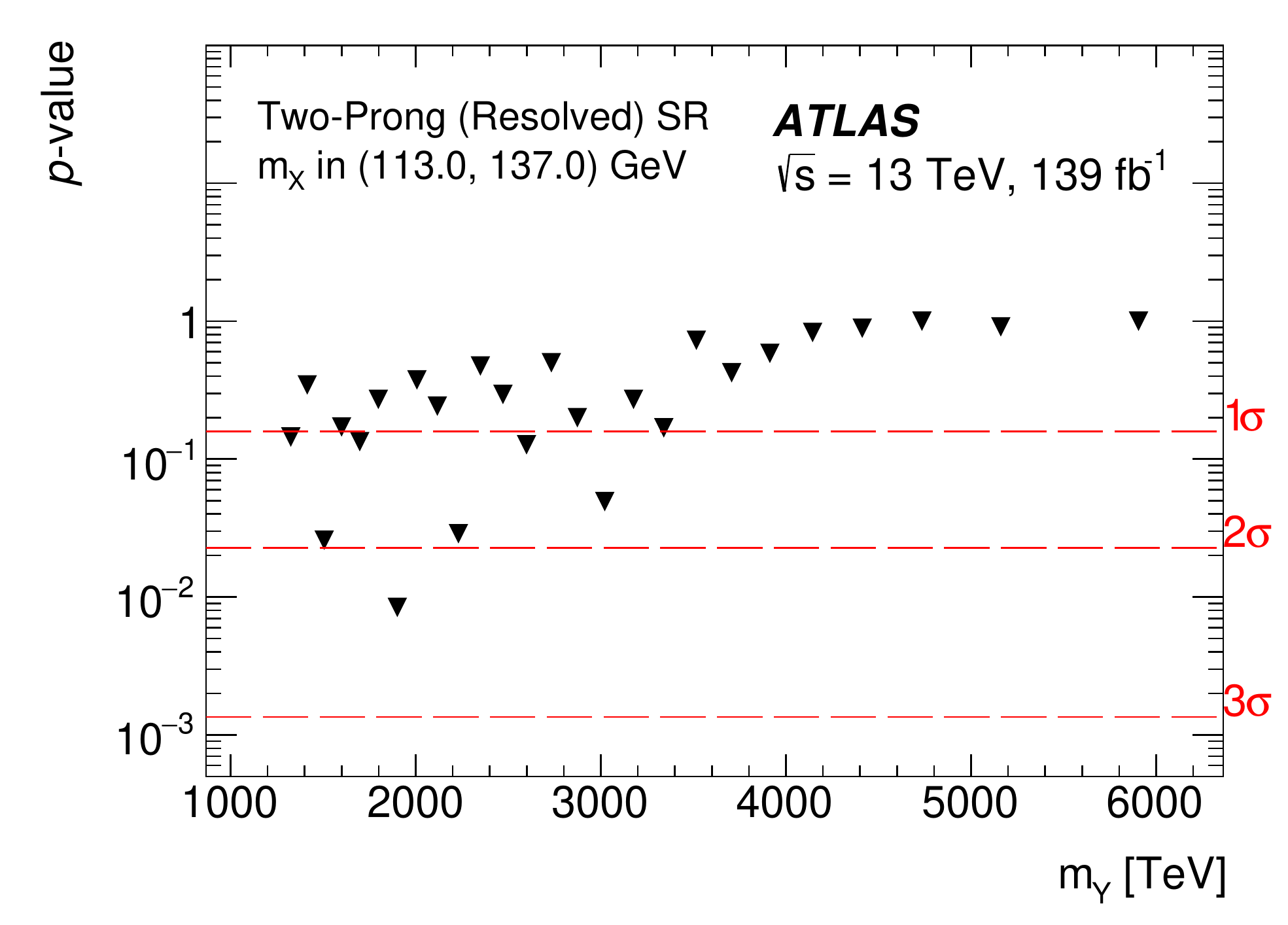}
\caption{\label{fig:results_excl}
The $p$-value per \my bin for both two-prong SRs, calculated using all systematic and statistical uncertainties of the background estimate. Two \mx bins are shown, [75.5, 95.5]  and [113.0, 137.0]~\GeV, which corresponds to a window containing the $W/Z$ and Higgs boson mass respectively. Events thus fall in (a) a merged $W/Z$ window, (b) a merged Higgs window, (c) a resolved $W/Z$ window, and (d) a resolved Higgs window. The background is determined by a background-only fit to the data with all statistical and systematic uncertainties included. In both \mx windows, the $p$-value has a high approximately constant value for the high $Y$-mass region of the resolved SR (bottom), as this phase-space region is far more likely to produce a highly boosted $J_X$ that falls in the merged SR selection.}
\end{figure}

Given the absence of significant excesses in the data, signal-plus-background fits are performed to determine the 95\% CL upper limit on the cross section of the \yxh process.
A summary of the expected and observed limits in the 2D grid of \yxh signals is given in Figure~\ref{fig:results_limits}, combining results from the merged and resolved regions.
A bilinear interpolation procedure is applied to provide results between fully simulated signal points.
The analysis is most sensitive in the very boosted regime, where the $Y$ mass is approximately an order of magnitude larger than the $X$ mass.
Sensitivity is lowest in the highly resolved regime, where the required large-$R$ $J$ reconstruction of the Higgs boson sculpts the signal efficiency to favor high-momentum $X$ particles.
The observed limits range from cross sections of 0.341~fb for the signal point ($\my = 5000$~\GeV, $\mx = 600$~\GeV), to 1.22~pb for the signal point ($\my = 2500$~\GeV, $\mx = 2000$~\GeV).

\begin{figure}[hbpt]
\centering
\includegraphics[width=0.45\textwidth]{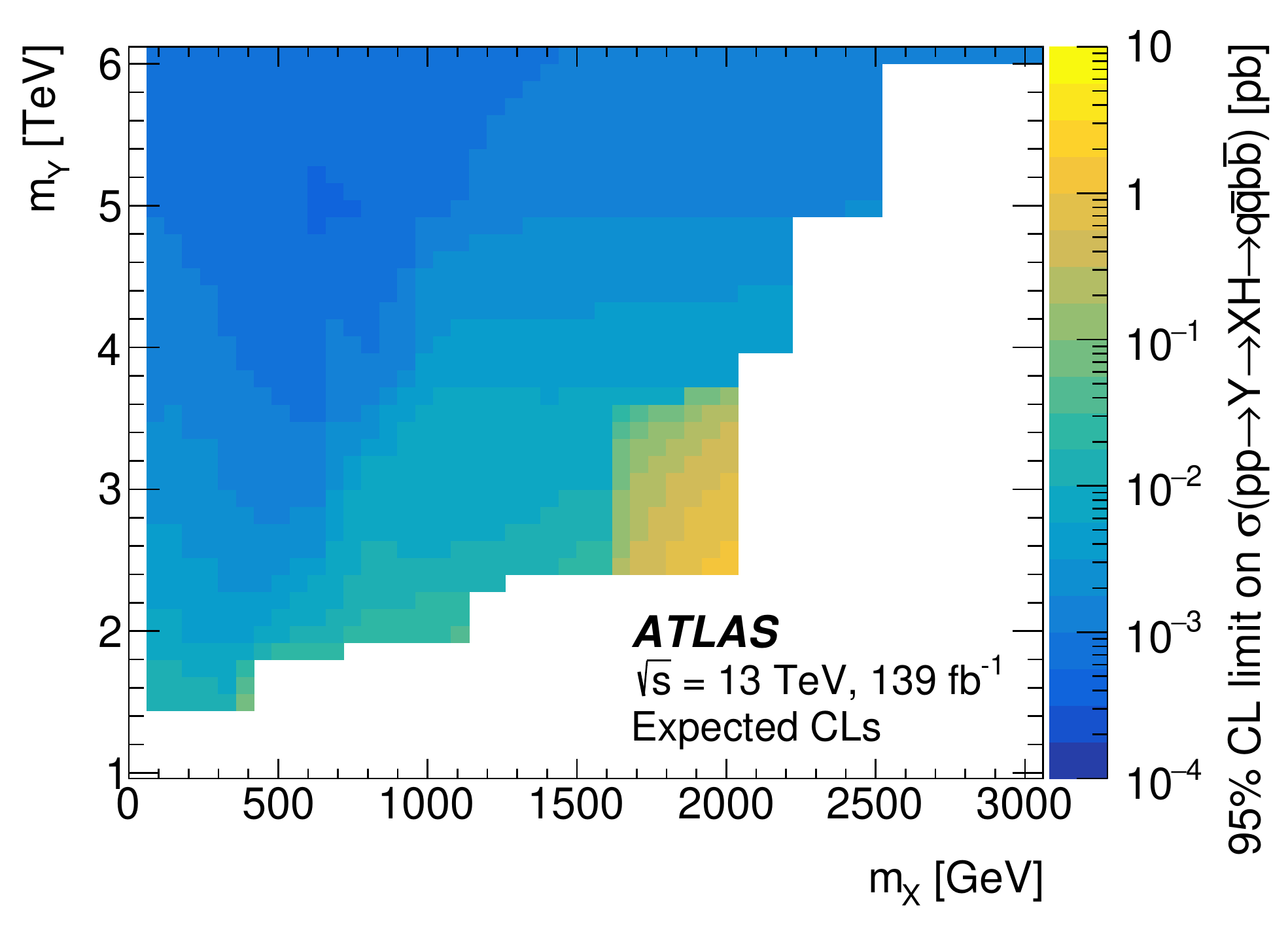}
\includegraphics[width=0.45\textwidth]{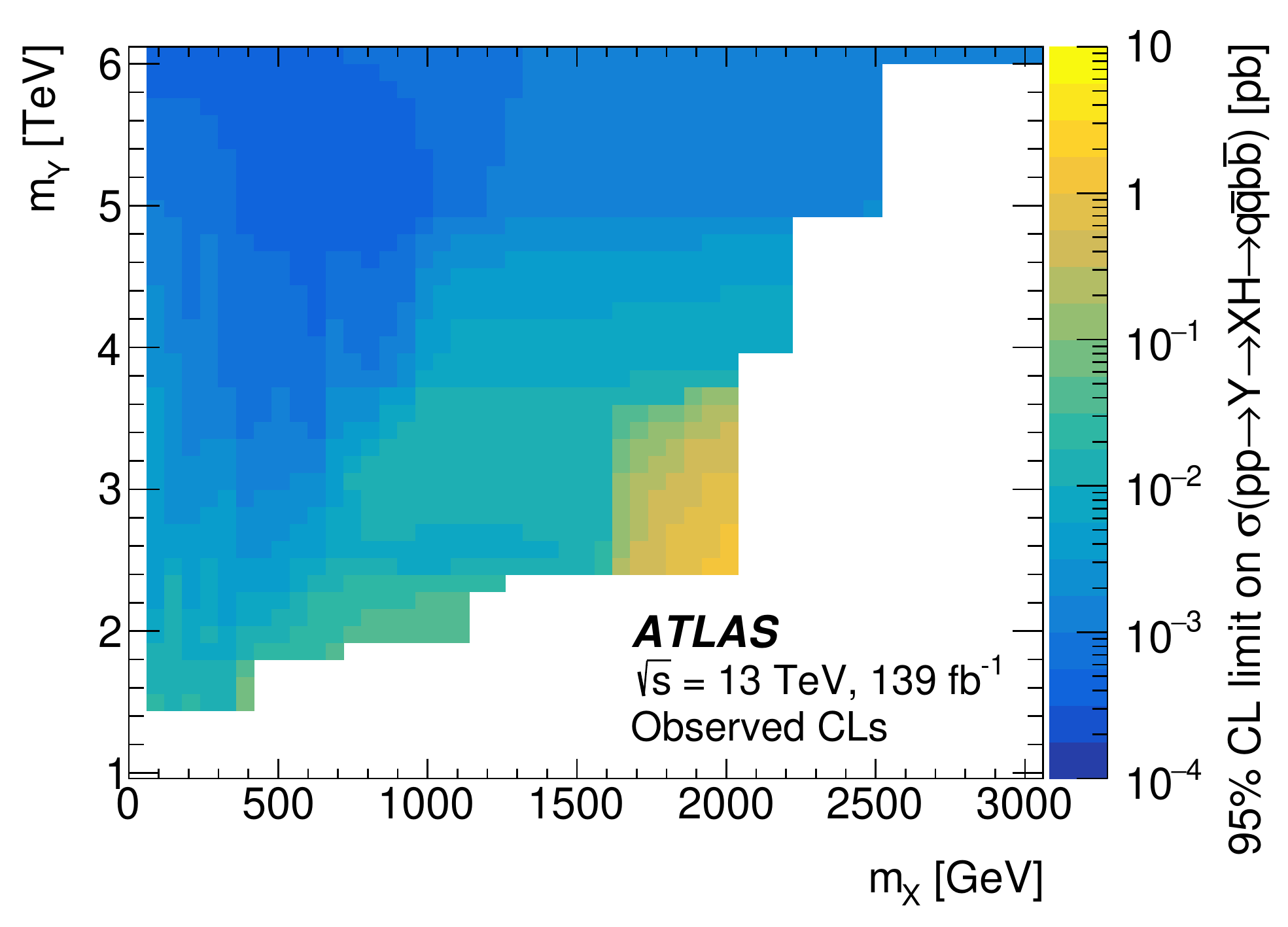}
\caption{\label{fig:results_limits}  The expected (left) and observed (right) 95\% CL limits on the cross section $\sigma(pp \rightarrow Y \rightarrow XH \rightarrow q\bar{q}b\bar{b}$) in pb in the two-dimensional space of \my~versus~\mx, obtained from a simultaneous fit of both the merged and resolved two-prong signal regions with all statistical and systematic uncertainties. A bilinear interpolation procedure is applied to provide results between fully simulated signal points. The observed limits range from 0.34~fb for the signal point ($\my = 5000$~\GeV, $\mx = 600$~\GeV) to 1.22~pb for the signal point ($\my = 2500$~\GeV, $\mx = 2000$~\GeV). The boundaries of the limit are defined by the simulated signal grid.}
\end{figure}
 
The data in the anomaly and two-prong SRs can be used to provide a benchmark comparison of sensitivity across the set of large-$R$ jet topologies considered for the \textit{X} decay, thereby providing a metric for assessing the level of signal-model-dependence in both regions.
The 95\% CL upper limits on the production cross section of several benchmark signals are calculated for all three signal region selections, by injecting signal into the data until the \textsc{BumpHunter} $p$-value exceeds a significance of 2$\sigma$.
Seven signal points are considered in this study, including three highly boosted \yxh points, one resolved \yxh point, and the three alternative jet topologies.
Because the systematic uncertainties in the signal efficiency of the Anomaly Score are not assessed, this comparison is performed using only statistical uncertainties and a post-fit background estimate in the limit calculation.
 
Since the merged region uses \dtrk and thus explicitly tags on the two-prong substructure of the \textit{X} candidate's large-$R$ jet in the generated \yxh grid, it is possible that this region will outperform the fully unsupervised approach for the \yxh signals.
Figure~\ref{fig:moneyplot} provides a summary of the limits obtained from the signal injection tests, comparing all three signal region selections for the seven benchmark signal points.
For points where the \textit{X} particle is highly boosted and thus the Anomaly Score is most sensitive, the upper limit on the cross section is approximately the same across the merged and anomaly SRs.
The \dtrk selection for this analysis is a general one; further optimization of the \dtrk selection would be expected to provide better performance than the Anomaly Score across two-prong simulated signals.
However, the signal-model-independent aspect of the anomaly detection used in the anomaly SR is more evident through improved limits on the alternative substructure signals.
Notably, for the dark jets the cross-section upper limit is nearly an order of magnitude lower than that provided by the two-prong regions, which underlines the particular strength of the model-independent approach for signals that are challenging to characterize with existing high-level variables.
 
\begin{figure}[hbpt]
\centering
\includegraphics[width=0.8\textwidth]{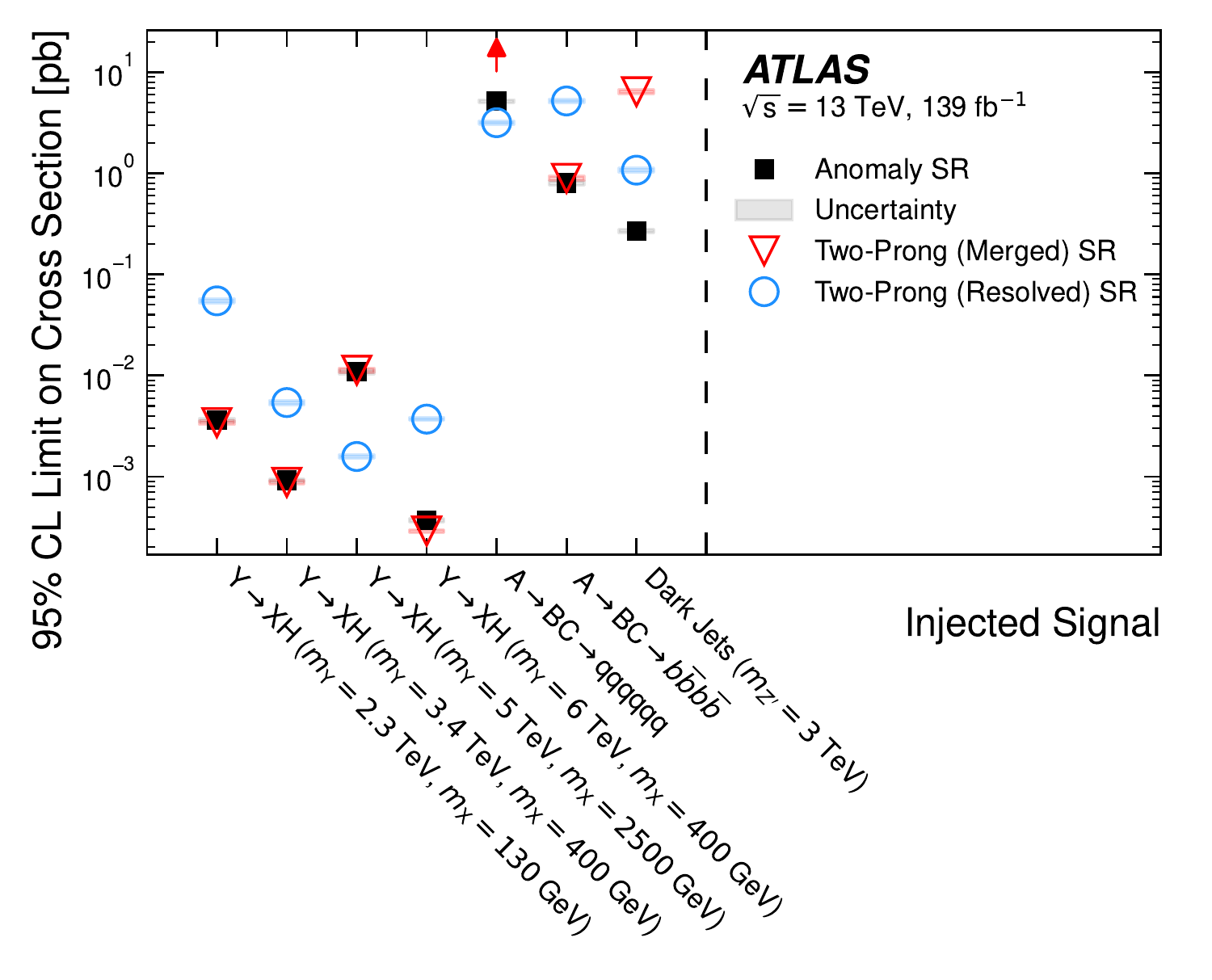}
\caption{\label{fig:moneyplot} The 95\% CL upper limit on the cross section for seven benchmark signal processes, comparing the anomaly, two-prong merged, and two-prong resolved signal region selections. The cross section is obtained by injecting signal into the observed data until 2$\sigma$ sensitivity is achieved as determined by the \textsc{BumpHunter} interval $p$-value. The background estimate is the result of a fit with all background uncertainties considered, while the \textsc{BumpHunter} algorithm incorporates only statistical uncertainties. The uncertainty band reflects the standard deviation of obtained cross section values for 100 trials. All signals are labeled by the masses of the BSM particles except the $A\rightarrow BC$ signals, where $m_A$ = 3 TeV, $m_B$ = 200 GeV, and $m_C$ = 400 GeV. Compared to the \dtrk-defined regions, the anomaly SR performs competitively for all signals considered, and considerably better for the dark jet signal, indicating a high degree of signal-model-agnostic sensitivity. The red upward-pointing arrow for $A\rightarrow BC\rightarrow qqqqqq$ indicates that the limit obtained from the two-pronged (merged) SR lies above the upper edge of the plot.}
\end{figure}
 
While these results cover regions of phase space that have not been studied directly by other searches, some analysis selections overlap with those of other recent ATLAS dijet resonance searches.
The \mx bin of [75.5, 95.5]~\GeV\ would be sensitive to a $VH$ resonance's hadronic final state, which is covered by a dedicated analysis using the same dataset~\cite{HDBS-2018-11}.
The present approach differs in the tagging techniques used for both the vector boson and Higgs boson, but provides a similar 95\% CL upper limit on the production cross section of a 3~\TeV\ resonance.
Due to its generality, the anomaly SR is expected to be sensitive to the same signatures as the weakly supervised dijet resonance search~\cite{HDBS-2018-59}, but a direct comparison is not provided due to the assumptions made here about the mass and decay of the  Higgs boson candidates.


\section{Conclusion}
\label{sec:conclusion}
 
A search is performed for a heavy new boson \textit{Y} decaying into a new particle \textit{X}  and a Standard Model Higgs boson in 139~\fb\ of 13~\TeV\ $pp$ collision data collected by the ATLAS detector during LHC \RunTwo.
The analysis focuses on a fully hadronic final state, where the \textit{X} particle and \textit{H} boson are boosted such that their daughter particles are collimated.
In the first application of fully unsupervised machine learning to an ATLAS search, a VRNN is trained on jets in data to define an anomaly detection SR, which selects the \textit{X} particle based solely on its substructural incompatibility with background jets.
Two supplementary SRs are designed to separately reconstruct merged and resolved decays of a nominal two-prong \textit{X} benchmark signal.
Sensitivity over the dominant multijet background is enhanced by additional machine-learning applications, namely a NN-based \Hbb tagger and a DNN-based reweighting to ensure good modeling.
No significant deviations of the data from the predicted background are observed.
The largest excess is found in the anomaly SR with a global significance of 1.43$\sigma$ when considering all \mx and \my bins, and is not found to be compatible with the expected signal shape.
Results are interpreted as 95\% confidence-level upper limits on the cross section $\sigma(pp \rightarrow Y \rightarrow XH \rightarrow q\bar{q}b\bar{b}$), across the two-dimensional space where \my is between 1.5 and 6~\TeV\ and \mx is between 65 and 3000~\GeV. The most stringent upper limit of 0.341~fb is obtained in the merged SR for the signal point ($\my = 5000$~\GeV, $\mx = 600$~\GeV), while the weakest upper limit of 1.22~pb is obtained for the highly resolved point ($\my = 2500$~\GeV, $\mx = 2000$~\GeV).


\section*{Acknowledgments}


We thank CERN for the very successful operation of the LHC, as well as the
support staff from our institutions without whom ATLAS could not be
operated efficiently.
 
We acknowledge the support of
ANPCyT, Argentina;
YerPhI, Armenia;
ARC, Australia;
BMWFW and FWF, Austria;
ANAS, Azerbaijan;
CNPq and FAPESP, Brazil;
NSERC, NRC and CFI, Canada;
CERN;
ANID, Chile;
CAS, MOST and NSFC, China;
Minciencias, Colombia;
MEYS CR, Czech Republic;
DNRF and DNSRC, Denmark;
IN2P3-CNRS and CEA-DRF/IRFU, France;
SRNSFG, Georgia;
BMBF, HGF and MPG, Germany;
GSRI, Greece;
RGC and Hong Kong SAR, China;
ISF and Benoziyo Center, Israel;
INFN, Italy;
MEXT and JSPS, Japan;
CNRST, Morocco;
NWO, Netherlands;
RCN, Norway;
MEiN, Poland;
FCT, Portugal;
MNE/IFA, Romania;
MESTD, Serbia;
MSSR, Slovakia;
ARRS and MIZ\v{S}, Slovenia;
DSI/NRF, South Africa;
MICINN, Spain;
SRC and Wallenberg Foundation, Sweden;
SERI, SNSF and Cantons of Bern and Geneva, Switzerland;
MOST, Taiwan;
TENMAK, T\"urkiye;
STFC, United Kingdom;
DOE and NSF, United States of America.
In addition, individual groups and members have received support from
BCKDF, CANARIE, Compute Canada and CRC, Canada;
PRIMUS 21/SCI/017 and UNCE SCI/013, Czech Republic;
COST, ERC, ERDF, Horizon 2020 and Marie Sk{\l}odowska-Curie Actions, European Union;
Investissements d'Avenir Labex, Investissements d'Avenir Idex and ANR, France;
DFG and AvH Foundation, Germany;
Herakleitos, Thales and Aristeia programmes co-financed by EU-ESF and the Greek NSRF, Greece;
BSF-NSF and MINERVA, Israel;
Norwegian Financial Mechanism 2014-2021, Norway;
NCN and NAWA, Poland;
La Caixa Banking Foundation, CERCA Programme Generalitat de Catalunya and PROMETEO and GenT Programmes Generalitat Valenciana, Spain;
G\"{o}ran Gustafssons Stiftelse, Sweden;
The Royal Society and Leverhulme Trust, United Kingdom.
 
The crucial computing support from all WLCG partners is acknowledged gratefully, in particular from CERN, the ATLAS Tier-1 facilities at TRIUMF (Canada), NDGF (Denmark, Norway, Sweden), CC-IN2P3 (France), KIT/GridKA (Germany), INFN-CNAF (Italy), NL-T1 (Netherlands), PIC (Spain), ASGC (Taiwan), RAL (UK) and BNL (USA), the Tier-2 facilities worldwide and large non-WLCG resource providers. Major contributors of computing resources are listed in Ref.~\cite{ATL-SOFT-PUB-2021-003}.


\clearpage

\printbibliography
 
\clearpage
\input{atlas_authlist}

\end{document}

%% file: atlas_authlist.tex
 
\begin{flushleft}
\hypersetup{urlcolor=black}
{\Large The ATLAS Collaboration}

\bigskip

\AtlasOrcid[0000-0002-6665-4934]{G.~Aad}$^\textrm{\scriptsize 101}$,
\AtlasOrcid[0000-0002-5888-2734]{B.~Abbott}$^\textrm{\scriptsize 119}$,
\AtlasOrcid[0000-0002-7248-3203]{D.C.~Abbott}$^\textrm{\scriptsize 102}$,
\AtlasOrcid[0000-0002-1002-1652]{K.~Abeling}$^\textrm{\scriptsize 55}$,
\AtlasOrcid[0000-0002-8496-9294]{S.H.~Abidi}$^\textrm{\scriptsize 29}$,
\AtlasOrcid[0000-0002-9987-2292]{A.~Aboulhorma}$^\textrm{\scriptsize 35e}$,
\AtlasOrcid[0000-0001-5329-6640]{H.~Abramowicz}$^\textrm{\scriptsize 150}$,
\AtlasOrcid[0000-0002-1599-2896]{H.~Abreu}$^\textrm{\scriptsize 149}$,
\AtlasOrcid[0000-0003-0403-3697]{Y.~Abulaiti}$^\textrm{\scriptsize 116}$,
\AtlasOrcid[0000-0003-0762-7204]{A.C.~Abusleme~Hoffman}$^\textrm{\scriptsize 136a}$
% ...
\bigskip
\\

% ...
$^{29}$Physics Department, Brookhaven National Laboratory, Upton NY; United States of America.\\
% ...
$^{35}$$^{(a)}$Facult\'e des Sciences Ain Chock, R\'eseau Universitaire de Physique des Hautes Energies - Universit\'e Hassan II, Casablanca;$^{(b)}$Facult\'{e} des Sciences, Universit\'{e} Ibn-Tofail, K\'{e}nitra;$^{(c)}$Facult\'e des Sciences Semlalia, Universit\'e Cadi Ayyad, LPHEA-Marrakech;$^{(d)}$LPMR, Facult\'e des Sciences, Universit\'e Mohamed Premier, Oujda;$^{(e)}$Facult\'e des sciences, Universit\'e Mohammed V, Rabat;$^{(f)}$Institute of Applied Physics, Mohammed VI Polytechnic University, Ben Guerir; Morocco.\\
% ...
$^{55}$II. Physikalisches Institut, Georg-August-Universit\"{a}t G\"ottingen, G\"ottingen; Germany.\\
% ...
$^{101}$CPPM, Aix-Marseille Universit\'e, CNRS/IN2P3, Marseille; France.\\
$^{102}$Department of Physics, University of Massachusetts, Amherst MA; United States of America.\\
% ...
$^{116}$Department of Physics, New York University, New York NY; United States of America.\\
% ...
$^{119}$Homer L. Dodge Department of Physics and Astronomy, University of Oklahoma, Norman OK; United States of America.\\
% ...
$^{136}$$^{(a)}$Departamento de F\'isica, Pontificia Universidad Cat\'olica de Chile, Santiago;$^{(b)}$Millennium Institute for Subatomic physics at high energy frontier (SAPHIR), Santiago;$^{(c)}$Instituto de Investigaci\'on Multidisciplinario en Ciencia y Tecnolog\'ia, y Departamento de F\'isica, Universidad de La Serena;$^{(d)}$Universidad Andres Bello, Department of Physics, Santiago;$^{(e)}$Instituto de Alta Investigaci\'on, Universidad de Tarapac\'a, Arica;$^{(f)}$Departamento de F\'isica, Universidad T\'ecnica Federico Santa Mar\'ia, Valpara\'iso; Chile.\\
% ...
$^{149}$Department of Physics, Technion, Israel Institute of Technology, Haifa; Israel.\\
$^{150}$Raymond and Beverly Sackler School of Physics and Astronomy, Tel Aviv University, Tel Aviv; Israel.\\
% ...
$^{a}$ Also Affiliated with an institute covered by a cooperation agreement with CERN.\\
$^{b}$ Also at An-Najah National University, Nablus; Palestine.\\
$^{c}$ Also at Borough of Manhattan Community College, City University of New York, New York NY; United States of America.\\
$^{d}$ Also at Bruno Kessler Foundation, Trento; Italy.\\
$^{e}$ Also at Center for High Energy Physics, Peking University; China.\\
$^{f}$ Also at Center for Interdisciplinary Research and Innovation (CIRI-AUTH), Thessaloniki ; Greece.\\
% ...

\end{flushleft}
